\documentclass[floatfix,showpacs,nofootinbib,twocolumn,showkeys]{revtex4-2}

\usepackage{graphicx}
\usepackage{etoolbox}
\usepackage{verbatim}
\usepackage{tensor}
\usepackage{url}
\usepackage{amsmath}
\usepackage{bm}
\usepackage{color}
\usepackage{hyperref}

\hypersetup{
    colorlinks=true,
    linkcolor=blue,
    filecolor=magenta,      
    urlcolor=blue,
    citecolor=blue
}

\usepackage{amsmath}
\DeclareMathOperator{\sign}{sign}

\newcommand{\feqh}{f_{{\rm eq},h}}
\newcommand{\feqhbar}{\bar{f}_{{\rm eq},h}}
\newcommand{\lambdapi}{\lambda_{\Pi} }

\definecolor{dartmouthgreen}{rgb}{0.05, 0.5, 0.06}

\begin{document}

\title{Maximum entropy kinetic matching conditions for heavy-ion collisions}
\date{\today}

\author{Derek~Everett}
\affiliation{Department of Physics, The Ohio State University, Columbus, OH 43210, USA}
\author{Chandrodoy Chattopadhyay}
\affiliation{Department of Physics, The Ohio State University, Columbus, OH 43210, USA}
\author{Ulrich Heinz}
\affiliation{Department of Physics, The Ohio State University, Columbus, OH 43210, USA}

\begin{abstract}
    Coupling hadronic kinetic theory models to fluid dynamics in phenomenological studies of heavy ion collisions requires a prescription for ``particlization''. Existing particlization models are based on implicit or explicit assumptions about the microscopic degrees of freedom that go beyond the information provided by the preceding fluid dynamical history. We propose an alternative prescription which uses only macroscopic information provided by the hydrodynamic output. This method follows directly from the connections between information theory and statistical mechanics.  
\end{abstract}

\maketitle

%
\section{Introduction}
%

When modeling heavy-ion collision dynamics macroscopically with fluid dynamics, the problem of particlization of the fluid near its decoupling into hadrons is a persistent source of theoretical model bias in the estimation of the material transport properties of the Quark-Gluon Plasma (QGP) liquid \cite{Teaney:2003kp, Dusling:2009df, Everett:2020yty}. Fluid dynamics provides only the hydrodynamic moments of the microscopic distributions of hadrons. For a fluid with conserved energy and momentum (but ignoring conserved charges) the stress-tensor $T^{\mu\nu}$ describes the energy and momentum fluxes. It is given by the second momentum-moment of the microscopic distribution,
\begin{equation}
\label{eqn:tmunu_match}
    T^{\mu\nu}(x) = \sum_h \frac{g_h}{(2\pi)^3} \int \frac{d^3p}{p_0} p^{\mu}p^{\nu}f_h(x;p),
\end{equation}
where the four-vectors $x$ and $p$ denote the space-time positions and particle momenta, $g_h$ is the spin-isospin degeneracy of hadronic species $h$, and $f_h(x;p)$ is the one-particle distribution function of species $h$. In local equilibrium the distribution function of each species is \textit{uniquely} specified by the macroscopic inverse temperature $\beta$ and the four-velocity of the fluid rest frame $u^{\mu}$. It is given by the J\"uttner distribution
\begin{equation}
\label{eqn:juttner}
    f_{{\rm eq}, h}(x;p) = \Bigl[\exp[\beta (u \cdot p)] - \theta \Bigr]^{-1},
\end{equation}
where $\theta = 1$, $0$, or $-1$ for particles obeying Bose-Einstein, Maxwell-Boltzmann, or Fermi-Dirac statistics, respectively. 

Out of local equilibrium however, there exist infinitely many microscopic distributions of hadron momenta and yields that reproduce the same hydrodynamic moments. Therefore, practitioners of hydrodynamic phenomenology often choose a particular ansatz for the microscopic physics when particlizing fluid cells in a hybrid hydrodynamic model of heavy-ion collisions. Commonly used ans\"atze include assumptions regarding the momentum-dependence of the viscous corrections to the local equilibrium distribution, such as the Grad (`14-moments') approach \cite{Israel:1976tn, Israel:1979wp, Monnai:2009ad}, and various approximations of an underlying kinetic theory based on the relativistic Boltzmann equation, e.g. the first-order Chapman-Enskog method with a collision term in Relaxation Time Approximation (RTA) \cite{ANDERSON1974466, Jaiswal:2014isa}. Both the Grad and Chapman-Enskog (CE) methods suppose that the microscopic distribution is split into two terms, the local-equilibrium distribution $\feqh$ and a dissipative correction $\delta f_h$, 
\begin{equation}
\label{eqn:linear_df}
    f_h(x;p) = f_{{\rm eq}, h}(x;p) + \delta f_h(x;p),
\end{equation}
and then solve their respective matching conditions to a finite (usually first) order in the viscous correction $\delta f_h(x;p)$.

In the Grad method one supposes that the viscous correction function may only have a quadratic dependence on the momenta,
\begin{equation}
\label{eqn:grad_df}
    \delta f_h(x;p) = f_{{\rm eq}, h} (1 - \theta f_{{\rm eq}, h})b_{\mu\nu} p^{\mu}p^{\nu},
\end{equation}
where $b_{\mu\nu}$ are coefficients which are fixed by the matching conditions (\ref{eqn:tmunu_match}) and thus linearly expressed in terms of the dissipative stresses, the bulk viscous pressure $\Pi$ and the shear stress tensor $\pi^{\mu\nu}$. Although at first order this does not involve a microscopic equation of motion \cite{Monnai:2009ad} (e.g. the Boltzmann equation) it does make a somewhat arbitrary assumption about the possible momentum-dependence of the viscous correction. This assumption can, however, be justified by relating the method to a systematic approximation of the Boltzmann equation in moments of $\delta f$ \cite{Denicol:2012cn}. The Chapman-Enskog series with the RTA collision term requires, at leading order, that the dissipative correction satisfies
\begin{equation}
\label{eqn:ce_df}
    \delta f_h =  - \frac{\tau_R}{u \cdot p} p^{\mu}\partial_{\mu}f_{\text{eq}} + \mathcal{O}\left(\partial^2 \right)
\end{equation}
where $\tau_R$ is the microscopic relaxation time. In practice, it is often assumed that the relaxation time $\tau_R = \tau_R(x)$ is species and momentum-independent, and this assumption is implied whenever we refer to the Chapman-Enskog RTA method in this manuscript.
Together with the matching conditions (\ref{eqn:tmunu_match}) this yields an ansatz for the viscous correction $\delta f_h$ which is linear in the dissipative stresses, but with different coefficients than Grad. 

It was realized early that the viscous corrections from the shear and bulk viscous stresses, motivated by various kinetic theories (the pion gas, perturbative QCD, RTA) had large effects on observables such as the elliptic flow \cite{Teaney:2003kp, Dusling:2009df}. Moreover, large bulk corrections in linearized approaches could overwhelm the equilibrium distribution resulting in the unphysical consequence $f_h = \feqh + \delta f_h <0$. This prompted the development of resummation methods designed such that the distribution functions are positive definite for all momenta \cite{Pratt:2010jt, Dusling:2011fd}.  

We note that many of these difficulties faced in the modeling of particlization are not unique to the field of heavy-ion collision phenomenology. Rather, this problem manifests whenever a fluid's coupling is not sufficiently strong, and its expansion sufficiently fast, such that the fluid decouples into particle degrees of freedom. Motivating a microscopic distribution given only knowledge of its macroscopic moments is a generic problem in kinetic theory and statistical mechanics. More broadly, motivating a unique probability distribution given only knowledge of its moments is a generic problem in information theory. The solution (to both) problems that provides the least-biased (maximally-entropic) distribution was given by Jaynes \cite{Jaynes:1957zza}. We follow these methods, tailoring them to the more unique concerns of particlization in heavy-ion collisions. These methods do not invoke a microscopic equation of motion, nor any {\it ad hoc} ans\"atze regarding the momentum dependence of the viscous corrections. The result is an expression for the microscopic distribution $f_h$ that depends non-linearly on the viscous stresses $\Pi$ and $\pi^{\mu\nu}$, is positive-definite, matches the entire stress-tensor and reduces to the linearized Chapman-Enskog in Relaxation Time Approximation form when the viscous stresses are weak.

Throughout this manuscript we will use natural units $\hbar = k_B = c = 1$ and the mostly-minus metric $g^{\mu\nu} = \rm{diag}(1, -1, -1, -1)$. Lorentz four-vector indices will be denoted by Greek letters while spatial three-vector indices are denoted by Latin letters. Contractions of Lorentz indices will sometimes be denoted by $(\cdot)$, e.g. $A_{\mu}B^{\mu}= A{\,\cdot\,}B$ or $p_\mu\pi^{\mu\nu}p_\nu=p{\,\cdot\,}\pi{\,\cdot\,}p$. 

%
\section{The maximum-entropy distribution}
%

%
%

The kinetic entropy density four-current $s^\mu(x)$ of a system of particles is given by
\begin{equation}
\label{eqn:entropy_current}
    s^{\mu}(x) = - \sum_h \frac{g_h}{(2\pi)^3}\int \frac{d^3p}{p_0} p^{\mu} \phi[f_h]
\end{equation}
where $f_h(x;p)$ is the one-particle distribution function, $p$ the momentum four-vector and $x$ the position four-vector. The function $\phi[f]$ depends on the quantum-statistical nature of the particles through the parameter $\theta$ defined in Eq.~(\ref{eqn:juttner}) and is defined by
\begin{equation}
\label{eqn:phi}
    \phi[f] \equiv f \ln (f) - \frac{1 + \theta f}{\theta} \ln(1 + \theta f).
\end{equation}

As is the case when performing particlization, let us suppose that some macroscopic theory (e.g. viscous hydrodynamics) provides us with the stress-energy tensor $T^{\mu\nu}$:
\begin{equation}
    T^{\mu\nu} = \epsilon u^{\mu}u^{\nu} - (p_{\rm eq} + \Pi)\Delta^{\mu\nu} + \pi^{\mu\nu}
\end{equation}
where $\epsilon$ is the energy density, $p_{\rm eq}$ the equilibrium pressure, $u^{\mu}$ the four-velocity vector, $\Pi$ the bulk viscous pressure and $\pi^{\mu\nu}$ the shear-viscous tensor. The energy-density and flow velocity are the eigenvalue and timelike eigenvector of the stress tensor 
\begin{equation}
    \epsilon u^{\mu} = u^{\nu} T^{\mu}_{\nu}.
\end{equation}
The equilibrium pressure can be related to the energy density by an equation of state $p_{\rm eq} = p_{\rm eq}(\epsilon)$, although it will not be necessary to do so in the method proposed in this manuscript. The spacelike projector $\Delta^{\mu\nu}$ is defined by
\begin{equation}
    \Delta^{\mu\nu} \equiv g^{\mu\nu} - u^{\mu}u^{\nu}.
\end{equation}
It is also convenient to define the symmetric and traceless projector 
\begin{equation}
\label{eqn:transv_traceless_proj}
    \Delta^{\mu\nu}_{\alpha\beta} \equiv \frac{1}{2}( \Delta^{\mu}_{\alpha}\Delta^{\nu}_{\beta} + \Delta^{\nu}_{\alpha}\Delta^{\mu}_{\beta} ) - \frac{1}{3}\Delta^{\mu\nu}\Delta_{\alpha\beta}.
\end{equation}
Then, the shear-stress tensor $\pi^{\mu\nu}$ is defined by
\begin{equation}
    \pi^{\mu\nu} = \Delta^{\mu\nu}_{\alpha\beta} T^{\alpha\beta},
\end{equation}
and the total isotropic pressure is defined by
\begin{equation}
    p_{\rm eq} + \Pi  = -\frac{1}{3}\Delta_{\mu\nu}T^{\mu\nu}.
\end{equation}

These definitions can be written as constraints on moments of the microscopic distributions $f_h(x;p)$: The energy matching condition requires 
\begin{equation}
\label{e}
    \epsilon = u_{\mu}T^{\mu\nu} u_\nu = \sum_h \frac{g_h}{(2\pi)^3} \int \frac{d^3p}{p_0} (u \cdot p)^2\, f_h(x;p).
\end{equation}
Matching the total isotropic pressure $P \equiv p_{\rm eq} + \Pi$ requires
\begin{equation}
\label{pPi}
    P = -\frac{1}{3}\Delta_{\mu\nu}T^{\mu\nu} = -\frac{1}{3} \Delta_{\mu\nu} \sum_h \frac{g_h}{(2\pi)^3} \int \frac{d^3p}{p_0} p^{\mu}p^{\nu} f_h(x;p).
\end{equation}
Finally, matching the shear-stress tensor requires
\begin{equation}
\label{eqn:shear_match}
    \pi^{\mu\nu} = \Delta^{\mu\nu}_{\alpha\beta} T^{\alpha\beta} = \Delta^{\mu\nu}_{\alpha\beta} \sum_h \frac{g_h}{(2\pi)^3} \int \frac{d^3p}{p_0} p^{\alpha}p^{\beta} f_h(x;p).
\end{equation}

Our approach here is to find the microscopic distribution $f_h(x;p)$ which maximizes the entropy density functional given only the ten components of $T^{\mu\nu}$ in Eq.~(\ref{eqn:tmunu_match}). We will do so using the canonical method of Lagrange multipliers \cite{Jaynes:1957zza}, without imposing any microscopic equation of motion. Our approach differs from Refs.~\cite{Calzetta:2010au, PeraltaRamos:2012xk} where, instead of the entropy, the {\it entropy production rate} was extremized. Computing that rate requires a microscopic approach -- in particular, one must specify the collision term. This incorporates additional information that we pretend not to possess --- in our work, we assume that all that is known is the energy-momentum tensor (\ref{eqn:tmunu_match}) resulting from the preceding hydrodynamic evolution of the fluid. In the absence of shear and bulk viscous stresses, our approach recovers the local equilibrium distribution (\ref{eqn:juttner}) --- the present work generalizes it to a maximum-entropy distribution for systems with non-zero viscous stresses.

After introducing Lagrange multipliers with the appropriate tensorial structure, the entropy density four-current can be written
\begin{eqnarray}
    s^{\mu}(x) =\ 
    && -\sum_h \frac{g_h}{(2\pi)^3} \int \frac{d^3p}{p_0} p^{\mu} \phi[f_h]
\nonumber\\
    && + \Lambda [\epsilon u^{\mu} - \sum_h \frac{g_h}{(2\pi)^3} \int \frac{d^3p}{p_0} p^{\mu} (u \cdot p) f_h]
\\
    && + \lambda^{\mu}[P + \frac{1}{3} \Delta_{\alpha\beta} \sum_h \frac{g_h}{(2\pi)^3} \int \frac{d^3p}{p_0} p^{\alpha}p^{\beta} f_h]
\nonumber\\\nonumber
    && + \gamma^{\mu}_{\langle\alpha\beta\rangle}[\pi^{\alpha\beta} - \Delta^{\alpha\beta}_{\rho\sigma} \sum_h \frac{g_h}{(2\pi)^3} \int \frac{d^3p}{p_0} p^{\rho}p^{\sigma} f_h ]. 
\end{eqnarray}
The Lagrange multipliers $\Lambda = \Lambda(x)$, $\lambda^\mu = \lambda^\mu(x)$ and $\gamma^\mu_{\alpha\beta} = \gamma^\mu_{\alpha\beta}(x)$ are all functions of spacetime $x$, although we do not explicitly write it for brevity of notation. In the last line we observed that $\pi^{\alpha\beta}=\Delta^{\alpha\beta}_{\rho\sigma}\pi^{\rho\sigma}\equiv \pi^{\langle\alpha\beta\rangle}$, where $\Delta^{\alpha\beta}_{\rho\sigma}$ is defined in Eq.~(\ref{eqn:transv_traceless_proj}), and used this to simplify the tensor structure of the Lagrange multiplier $\gamma^{\mu}_{\alpha\beta}$.

We seek the distribution $f_h$ which maximizes the entropy density in the local rest frame $u \cdot s$:
\begin{equation}
\label{eqn:maximize_entropy}
    \frac{\delta (u \cdot s)}{\delta f_h} = 0.
\end{equation}
It follows that
\begin{eqnarray}
   &&\ln \left[ \frac{f_h}{1+\theta f_h} \right] =  
\\\nonumber
   &&-\Lambda (u \cdot p) + \frac{u \cdot \lambda}{u \cdot p}\Delta_{\alpha\beta}p^{\alpha}p^{\beta} - \frac{1}{u \cdot p}u_{\mu}\gamma^{\mu}_{\alpha\beta} \Delta^{\alpha\beta}_{\rho\sigma} p^{\rho} p^{\sigma}
\end{eqnarray}
and, after simplification and exponentiation, yields the main result:
\begin{eqnarray} 
\label{f_h}
   &&f_h(x,p) = 
\\\nonumber
   &&\Bigl[ \exp\Bigl(\Lambda\bigl(u \cdot p\bigr) - \frac{u \cdot \lambda}{u \cdot p} p_{\langle\alpha\rangle} p^{\langle\alpha\rangle} + \frac{u_{\mu}\gamma^{\mu}_{\langle\alpha\beta\rangle}}{u \cdot p} p^{\langle\alpha} p^{\beta\rangle} \Bigr)  - \theta \Bigr]^{-1}.
\end{eqnarray}
We refer to Eq.~(\ref{f_h}) as the maximum-entropy (ME) distribution.
Here $p^{\langle\alpha\rangle}=\Delta^{\alpha\beta}p_\beta$ denotes the spatial components of $p_\mu$ in the local rest frame (LRF), defined by $u^{\mu}_{\rm LRF} = (1, \mathbf{0})$. In that frame the maximum-entropy distribution is given by
\begin{equation}
\label{f_h_LRF}
   f_h^{\text{LRF}} = \Bigl[ \exp\Bigl(\Lambda p_0 + \frac{\lambdapi}{p_0} \bm{p}^2
   + \frac{\gamma^{0}_{ij} p^i p^j}{p_0} \Bigr) - \theta \Bigr]^{-1},
\end{equation}
where $p_0=\sqrt{m^2{+}\bm{p}^2}$, $\gamma^{0}_{ij}$ is traceless and symmetric in the spatial indices $(ij)$, and we have defined $\lambdapi \equiv u \cdot \lambda$. This equation can be rewritten in a form which bears resemblance to previous particlization ans\"atze which have been studied in the past \cite{Pratt:2010jt},
\begin{equation}
\label{eq22}
    f_h(x;\bm{p}) = \Bigl[ e^{\Lambda p_0 }\, \exp\Bigl(\Lambda_{ij} \frac{p^{i}p^{j}}{p_0}\Bigr) - \theta \Bigr]^{-1},
\end{equation}
where the linear-transformation operator $\Lambda_{ij}$ (which acts on the spatial momenta in the LRF) is defined by
\begin{equation}
    \Lambda_{ij}(x) \equiv 
    \lambdapi(x) \delta_{ij} + \gamma^{0}_{ij}(x).
\end{equation}

This maximum-entropy distribution, in particular the tensor structure of the transformation $\Lambda_{ij}$, indeed bears striking resemblance to the so-called ``modified equilibrium'' distributions \cite{Pratt:2010jt,  McNelis:2019auj}; however, we find the coefficients in $\Lambda_{ij}$ are different. Eq.~(\ref{eq22}) also shares some structural similarities with ``anisotropic equilibrium distribution'' functions 
\cite{Florkowski:2013lza, Alqahtani:2017mhy, Alqahtani:2017tnq, Nopoush:2019vqc}; again, a closer comparison reveals differences.  We show in section \ref{sec:linearized_me} that if one works to linear order in the dissipative stresses $\pi^{\mu\nu}$ and $\Pi$, the maximum-entropy prescription matches exactly the Chapman-Enskog RTA method. This feature is also shared by particular modified equilibrium approaches \cite{Pratt:2010jt, McNelis:2019auj}, meaning the relation between the maximum-entropy, modified equilibrium and Chapman-Enskog RTA approaches is exact at linear order in the dissipative stresses. However, once second-order and higher terms have been included, this equivalence is broken and all three prescriptions differ. 

We note that for Maxwell-Boltzmann particles ($\theta=0$) the non-equilibrium entropy density of our system admits a thermodynamic expression in terms of the hydrodynamic fields and their conjugate variables. In this case, the entropy density in the local rest frame $s$ is given by
\begin{equation}
\label{s_MB}
     s = \Lambda \epsilon + \Lambda_{ij} P^{ij} + n,
\end{equation}
where $n$ is the particle density and we've defined the pressure tensor $P^{ij}$ by
\begin{equation}
    P^{ij} \equiv \sum_h \frac{g_h}{(2\pi)^3} \int \frac{d^3p}{p_0} p^ip^j f_h, 
\end{equation}
which contains both the equilibrium pressure $p_{\rm eq}\delta^{ij}$ and viscous corrections.
A more general thermodynamic relation which holds for Bose-Einstein and Fermi-Dirac statistics as well can be defined by introducing a generating function $\mathcal{Z}$. 
The expression is given by
\begin{equation}\label{s_gen}
    s = \Lambda \epsilon + \Lambda_{ij} P^{ij} + {\cal Z}(\Lambda, \Lambda_{ij}),
\end{equation}
where the generating function $\mathcal{Z}$
\begin{equation}
\label{genfunc0}
    {\cal Z} \equiv \sum_{h} \frac{g_h}{(2\pi)^3} \int d^3p \, \frac{1}{\theta} \, \ln{\left(1 + \theta \, f_{h} \right)}
\end{equation}
%
is defined such that its derivatives generate the hydrodynamic fields,
\begin{equation}
\label{derivatives}
    \frac{\partial {\cal Z}}{\partial \Lambda} = - \epsilon, \qquad \frac{\partial {\cal Z}}{\partial \Lambda_{ij}} = - P^{ij}.
\end{equation}

%

Our distribution function is expressed in terms of seven unknown Lagrange multipliers $\Lambda$ and $\Lambda_{ij}$ (or, covariantly, $\Lambda$, $u\cdot\lambda$, and $\gamma_{\langle\mu\nu\rangle}\equiv u_\alpha\gamma_{\langle\mu\nu\rangle}^\alpha$) which must be chosen to match the energy density, total isotropic pressure and shear-stress tensor, respectively. To solve for these coefficients, we write down the seven required matching conditions (\ref{e})-(\ref{eqn:shear_match}) in terms of LRF momenta and components as follows:
\begin{eqnarray}
\label{eqn:ematch}
    &&\epsilon = \sum_h \frac{g_h}{(2\pi)^3} \int d^3p\, p_0\, \Bigl[ e^{\Lambda p_0 }\, \exp\Bigl(\Lambda_{ij} \frac{p^{i}p^{j}}{p_0}\Bigr) - \theta \Bigr]^{-1}\!\!\!\!,\quad
\\
\label{Pimatch}
    &&P = \sum_h \frac{g_h}{(2\pi)^3} \int \frac{d^3p}{p_0}\, \frac{\bm{p}^2}{3}\, \Bigl[ e^{\Lambda p_0}\, \exp\Bigl(\Lambda_{ij} \frac{p^{i}p^{j}}{p_0}\Bigr) - \theta \Bigr]^{-1}\!\!\!\!,\qquad
\\
\label{pimatch}
    &&\pi^{ij} = \sum_h \frac{g_h}{(2\pi)^3} \int \frac{d^3p}{p_0} \Bigl(p^i p^j - {\textstyle\frac{1}{3}} \bm{p}^2 \delta^{ij}\Bigr)\, 
\nonumber\\
    &&\hspace*{3.3cm} \times\Bigl[ e^{\Lambda p_0}\, \exp\Bigl(\Lambda_{kl} \frac{p^{k}p^{l}}{p_0}\Bigr) - \theta \Bigr]^{-1}\!\!\!.
\end{eqnarray}
Note that, since we only introduced a single Lagrange multiplier for each constraint, i.e. we maximized the LRF entropy density only subject to the information given to us directly through the energy momentum tensor, the maximum entropy distribution (\ref{f_h}) automatically distributes the shear and bulk viscous flows ``democratically'' \cite{Molnar:2014fva} across the hadron species $h$. 

%
\section{Linearizing the maximum entropy distribution}
\label{sec:linearized_me}
%

In this section we solve for the Lagrange multipliers in the limit of small dissipative flows. We calculate the maximum entropy distribution self-consistently to leading order in the viscous stresses $\Pi$ and $\pi^{\mu\nu}$, and find that it reproduces exactly the Chapman-Enskog distribution in the Relaxation-Time Approximation. We begin by noting that for vanishing viscous stresses (ideal fluids) the constraint (\ref{eqn:ematch}) is solved by setting $\Lambda_{ij}\equiv 0$ and $\Lambda=\beta$, the equilibrium inverse-temperature. The leading order expressions can therefore be obtained by expanding the maximum entropy distribution (\ref{f_h}) to linear order in $\Lambda_{ij}$ and the difference $\Lambda{\,-\,}\beta$. For simplicity we will assume Maxwell-Boltzmann statistics ($\theta=0$) throughout this section. We also introduce the compact notation
\begin{equation}
    \int_p (\cdots) \equiv \sum_h \frac{g_h}{(2\pi)^3} \int \frac{d^3p}{p^0} (\cdots),
\end{equation}
where $p^0=\sqrt{\bm{p}^2+m_h^2}$. (Note that the shorthand $\int_p$ includes a sum over species $h$.) 

Let us consider Eq.~(\ref{f_h}) and for notational convenience define the traceless and purely spatial (in the LRF) rank two-tensor $\gamma_{\alpha\beta}(x) = \gamma_{\langle\alpha\beta\rangle}(x) \equiv u_\mu(x) \gamma^{\mu}_{\langle\alpha \beta\rangle}(x)$. The maximum-entropy distribution (\ref{f_h}) is then expressed as
\begin{equation}
    f_h = \exp \left(- \Lambda\, u{\,\cdot\,}p + \lambda_\Pi\frac{p_{\langle\mu\rangle} p^{\langle\mu\rangle}}{u \cdot p}  - \gamma_{\langle\mu\nu\rangle} \frac{p^{\langle\mu} p^{\nu\rangle}}{u \cdot p} \right).
\end{equation}
This distribution depends on the hadron species $h$ only through the mass dependence of the on-shell energy.
The local-equilibrium distribution is given by 
\begin{equation}
    \feqh = e^{-\beta (u \cdot p)}.
\end{equation}

Let us now consider the relation between $\Lambda$, $\lambda_\Pi$, $\gamma_{\langle\mu\nu\rangle}$ and $\beta$ provided by the energy matching condition Eq.~(\ref{eqn:ematch}) discussed at the end of the preceding section and expand it around $\lambda_\Pi=\gamma^{\langle\mu\nu\rangle}=0$:
\begin{equation}
    \Lambda(\beta,\lambda_\Pi, \gamma^{\langle\mu\nu\rangle}) = \Lambda(\beta, 0 , 0) + c_\lambda \lambda_\Pi + c_{\mu\nu} \gamma^{\langle\mu\nu\rangle} + \cdots.
\end{equation}
Here $\Lambda (\beta, 0, 0) = \beta$, the coefficients $c_\lambda$ and $c_{\mu\nu}$ are in general functions of $\beta$, and the ellipses denote neglected terms of second or higher order in $\lambdapi$ and $\gamma^{\mu\nu}$. To linear order in the Lagrange multipliers the distribution function $f_h$ can thus be written 
\begin{eqnarray} 
\label{f_CElike}
    &&f_h \approx e^{-\beta (u \cdot p)} 
    \Bigl[1 - \bigl(c_\lambda \lambda_\Pi + c_{\mu\nu} \gamma^{\langle\mu\nu\rangle}\bigr) (u{\,\cdot\,}p)  
\nonumber\\
    &&\hspace*{2.4cm} 
    +\, \lambda_\Pi\frac{p_{\langle\mu\rangle} p^{\langle\mu\rangle}}{u \cdot p}  - \gamma^{\langle\mu\nu\rangle} \frac{p_{\langle\mu} p_{\nu\rangle}}{u\cdot p}\Bigr].
\end{eqnarray}
The matching conditions for the shear and bulk stresses can be compactly written as
\begin{equation}
    \int_p p^\mu p^\nu\, \delta f_h = - \Pi \Delta^{\mu\nu} + \pi^{\mu\nu},
\end{equation}
where $\delta f_h$ is defined in Eq.~(\ref{eqn:linear_df}). Using the linear\-ized form (\ref{f_CElike}) and introducing the shorthand $c_\lambda \lambdapi + c_{\mu\nu} \gamma^{\langle\mu\nu\rangle} \equiv {\cal C}$ this becomes
\begin{widetext}
\begin{equation} 
\label{CE_matching}
    - {\cal C} \int_p (u \cdot p)\, p^\mu p^\nu\, \feqh +  \lambda_\Pi \int_p p^\mu p^\nu \frac{p_{\langle\alpha\rangle} p^{\langle\alpha\rangle}}{u \cdot p} \feqh - \gamma_{\langle\alpha\beta\rangle} \int_p    \frac{p^\mu p^\nu p^\alpha p^\beta}{u\cdot p} \feqh
    = - \Pi \Delta^{\mu\nu} + \pi^{\mu\nu}.
\end{equation}
To calculate the left hand side we use the following tensor decomposition for $n=4$:
\begin{align}\label{tensor_decom}
\nonumber
    \int_p \frac{p^{\mu_1} p^{\mu_2} \cdots p^{\mu_n}}{(u \cdot p)^r} \feqh = I_{(n,0)}^{r} u^{\mu_1} u^{\mu_2} \cdots u^{\mu_n}
    + I_{(n,1)}^{r} \Big[ \Delta^{\mu_1 \mu_2} u^{\mu_3} \cdots u^{\mu_n} + \Delta^{\mu_1 \mu_3} u^{\mu_2} u^{\mu_4} \cdots u^{\mu_n} +
    \mathrm{permutations} \Big] + \cdots
\end{align}
\end{widetext}
where the moments $I_{(n,q)}^{r}$ are defined as
\begin{equation}
    I_{(n,q)}^{r} \equiv \frac{1}{(2q + 1)!!} \int_p (u \cdot p)^{n-2q-r} \, \bigl(p^{\langle\alpha\rangle} p_{\langle\alpha\rangle}\bigr)^{q} \, \feqh.
\end{equation}
Note that the authors of Ref.~\cite{Jaiswal:2014isa} used a similar definition, albeit for a single hadron species. Labeling the moments used in \cite{Jaiswal:2014isa} as $I_{(n,q)}^{r,h}$ for a given species $h$, the two definitions are simply related by $I_{(n,q)}^{r} = \sum_{h} I_{(n,q)}^{r,h}$.

After tensor decomposition we find
\begin{align}
    &\int_p (u \cdot p)\, p^\mu p^\nu\, \feqh 
    = I_{(2,0)}^{-1} u^\mu u^\nu + I_{(2,1)}^{-1} \Delta^{\mu\nu},
\\
    &\int_p p^\mu p^\nu \frac{p_{\langle\alpha\rangle} p^{\langle\alpha\rangle}}{u \cdot p} \feqh
    = 3 I_{(4,1)}^{1} u^\mu u^\nu + 5 I_{(4,2)}^{1} \Delta^{\mu\nu},
\\
    &\gamma_{\langle\alpha\beta\rangle} \int_p \frac{p^\mu p^\nu p^\alpha p^\beta}{u\cdot p} \feqh
    = 2 I_{(4,2)}^{1} \, \gamma^{\langle\mu\nu\rangle}. 
\end{align}
In the last line we used that $\gamma^{\langle\mu\nu\rangle}$ is symmetric, traceless and transverse to the flow velocity $u^\mu$. Substituting this back into Eq.~(\ref{CE_matching}) yields
\begin{eqnarray}
    - \Pi \Delta^{\mu\nu} &{+}& \pi^{\langle\mu\nu\rangle}
    =\bigl(- {\cal C} I_{(2,0)}^{-1} + 3 \lambdapi I_{(4,1)}^{1} \bigr) u^{\mu} u^{\nu} 
\\\nonumber
    &{+}&\, \bigl(- {\cal C} I_{(2,1)}^{-1} + 5 \lambdapi I_{(4,2)}^{1} \bigr) \Delta^{\mu\nu} 
    - 2 I_{(4,2)}^{1} \gamma^{\langle\mu\nu\rangle}.
\end{eqnarray}
Using the mutual orthogonality of the tensors $u^\mu u^\nu$, $\Delta^{\mu\nu}$ and $\pi^{\langle\mu\nu\rangle}$ we find the following three relations:
\begin{subequations}
\begin{align}
- {\cal C} I_{(2,0)}^{-1} + 3 \lambda_\Pi I_{(4,1)}^{1} & = 0, \\
- {\cal C} I_{(2,1)}^{-1} + 5 \lambda_\Pi I_{(4,2)}^{1} & = -\Pi, \\
- 2 I_{(4,2)}^{1} \gamma^{\langle\mu\nu\rangle} = \pi^{\langle\mu\nu\rangle}.
\end{align}
\end{subequations}
With the help of Eqs.~(8, 11, 12, 22) in Ref.~\cite{Jaiswal:2014isa} for a single hadron species, remembering that here $I_{(n,q)}^{r} = \sum_{h} I_{(n,q)}^{r,h}$ as well as $\beta_\pi =\sum_h \beta_{\pi,h}$ (where $\beta_{\pi,h} \equiv \beta I_{(4,2)}^{1,h}$ was defined in \cite{Jaiswal:2014isa}), we find
\begin{subequations}
\begin{align}
    &I_{(2,0)}^{-1} = I_{(3,0)}^{0}, \quad
     I_{(2,1)}^{-1}  = I_{(3,1)}^{0}, \quad
     I_{(4,1)}^{1}   = I_{(3,1)}^{0},  
\\
    &I_{(4,2)}^{1} \equiv \beta_\pi/\beta, \quad 
     I_{(3,1)}^{0} = - (\epsilon + p)/\beta, 
\\
    &I_{(3,1)}^{0} / I_{(3,0)}^{0} = - dp_{eq}/d\epsilon \equiv - c_s^2.
\end{align}
\end{subequations}
In the last equation we used $dp_{eq} = \sum_{h} d p_{eq,h}$ and $d\epsilon = \sum_{h} d\epsilon_{h}$, as well as $dp_{eq,h} =  I_{(3,1)}^{0,h} d\beta$ and $d \epsilon_{h} = - I_{(3,0)}^{0,h} d\beta$, such that $c_s^2 = - I_{(3,1)}^{0}/I_{(3,0)}^{0}$.

Putting everything together we find the following relations between the Lagrange multipliers, coefficients and dissipative stresses:
\begin{subequations}
\label{eqn:CE_RTA_coeffs}
\begin{align}
    {\cal C} & = - 3 \lambda_\Pi c_s^2, \quad
    \gamma^{\mu\nu} = - \beta \pi^{\mu\nu}/(2 \beta_\pi), 
\\
    \lambda_\Pi &= - \frac{\beta \, \Pi}{3 [ \frac{5}{3} \beta_\pi - (\epsilon + p) c_s^2]} \equiv - \frac{\beta \, \Pi}{3 \beta_\Pi},
\end{align}
\end{subequations}
where, similar to the corresponding definition in \cite{Jaiswal:2014isa},
\begin{equation}
    \beta_\Pi = \frac{5}{3}\beta I_{(4,2)}^1 - (\epsilon{+}p_{\rm eq}) c_s^2.
\end{equation}
Using these results in Eq.~(\ref{f_CElike}) yields for the linearized maximum-entropy viscous correction $\delta f_h$ the expression 
\begin{widetext}
\begin{equation}
\label{eqn:delta_f_CE_RTA}
    \frac{\delta f_h}{\feqh} = \Bigl(3 c_s^2 (u \cdot p) 
    + \frac{p_{\langle\alpha\rangle} p^{\langle\alpha\rangle}}{u \cdot p}\Bigr)\lambda_\Pi   
    - \frac{p_\mu p_\nu \gamma^{\langle\mu\nu\rangle}}{u \cdot p} 
    = \frac{\beta}{3\beta_\Pi} \Bigl((1{-}3 c_s^2)(u\cdot p)^2-m_{h}^2\Bigr) \frac{\Pi}{u{\,\cdot\,}p} + \frac{\beta}{2\beta_\pi} 
    \frac{p_\mu p_\nu \pi^{\mu\nu}}{u\cdot p}.
\end{equation}
For a single hadron species this matches exactly with Eq.~(27) in Ref.~\cite{Jaiswal:2014isa} which was derived by solving the first-order Chapman-Enskog correction with a relaxation-time approximation collision kernel.
\end{widetext}

%
\section{Matching without shear stress}
\label{sec:no_shear}
%

We now consider the case in which we match an energy-momentum tensor with vanishing shear-stress, $\pi^{\mu\nu} = 0$. This implies that the associated Lagrange multiplier $\gamma_{\langle\mu\nu\rangle}=0$ and that the distribution (\ref{f_h}) is isotropic in the LRF. Using LRF momenta (i.e. $p_0=u\cdot p$) to evaluate the matching integrals (\ref{eqn:ematch})--(\ref{Pimatch}) we obtain
\begin{widetext}
\begin{eqnarray}
\label{eqn:energy_integral}
    \epsilon &=& \frac{4\pi}{(2\pi)^3} \sum_h g_h \int_{m_h}^{\infty} dp_0\, p_0^2 \sqrt{p_0^2{-}m_h^2} \Bigl [ \exp\Bigl(\Lambda p_0 + \frac{\lambda_{\Pi}}{p_0}(p_0^2{-}m_h^2)\Bigr) - \theta \Bigr]^{-1},
\\
\label{eqn:pressure_integral}
    P = p_{\rm eq} + \Pi &=& \frac{4\pi}{(2\pi)^3} \frac{1}{3} \sum_h g_h \int_{m_h}^{\infty} dp_0\,(p_0^2{-}m_h^2)^{3/2} \Bigl [ \exp\Bigl(\Lambda p_0 + \frac{\lambda_{\Pi}}{p_0}(p_0^2{-}m_h^2)\Bigr) - \theta \Bigr]^{-1}.
\end{eqnarray}
\end{widetext}
Examining the expression for the distribution function 
\begin{equation}
   f = \left[\exp\Bigl(\Lambda p_0 + \frac{\lambda_{\Pi}}{p_0}(p_0^2{-}m_h^2)\Bigr) - \theta \right]^{-1},
\end{equation}
where $\Lambda > 0$, we note that existence of a solution requires $\lambdapi \geq -\Lambda$. Since large values of $|\lambdapi|$ signal large bulk viscous stresses, we restrict our solution to the case $|\lambdapi| \leq |\Lambda|$.\footnote{%
    Technically we ensure this by, instead of $\lambdapi$, using $\rho \equiv \lambdapi /  \Lambda$ and restricting $\rho$ to the range $[-1, 1]$.}

To solve Eqs.~(\ref{eqn:energy_integral},\ref{eqn:pressure_integral}) numerically we use a realistic hadron resonance gas (HRG) which includes all resonances that can be propagated in the \texttt{UrQMD} \cite{Bass:1998ca, Bleicher:1999xi} hadronic afterburner. We solve these coupled equations for $\Lambda$ and $\lambdapi$ as follows:
First, we choose a regular grid of values for both $\Lambda$ and $\lambdapi$. Then, for each pair of values $(\Lambda_i, \lambda_{\Pi,j})$ we evaluate the integrals in Eqs.~(\ref{eqn:energy_integral}) and (\ref{eqn:pressure_integral}) by numerical quadrature. This yields a grid of values $\epsilon(\Lambda_i, \lambda_{\Pi,j})$ and $P(\Lambda_i, \lambda_{\Pi,j})$. These grids are then interpolated with splines to obtain smooth approximations $\hat{e}(\Lambda, \lambdapi)$ and $\hat{P}(\Lambda, \lambdapi)$. Finally, given known values of energy density and isotropic pressure, $\Lambda$ and $\lambdapi$ can be found using a two-dimensional root finding routine. 

\begin{figure}
\includegraphics[width=8cm]{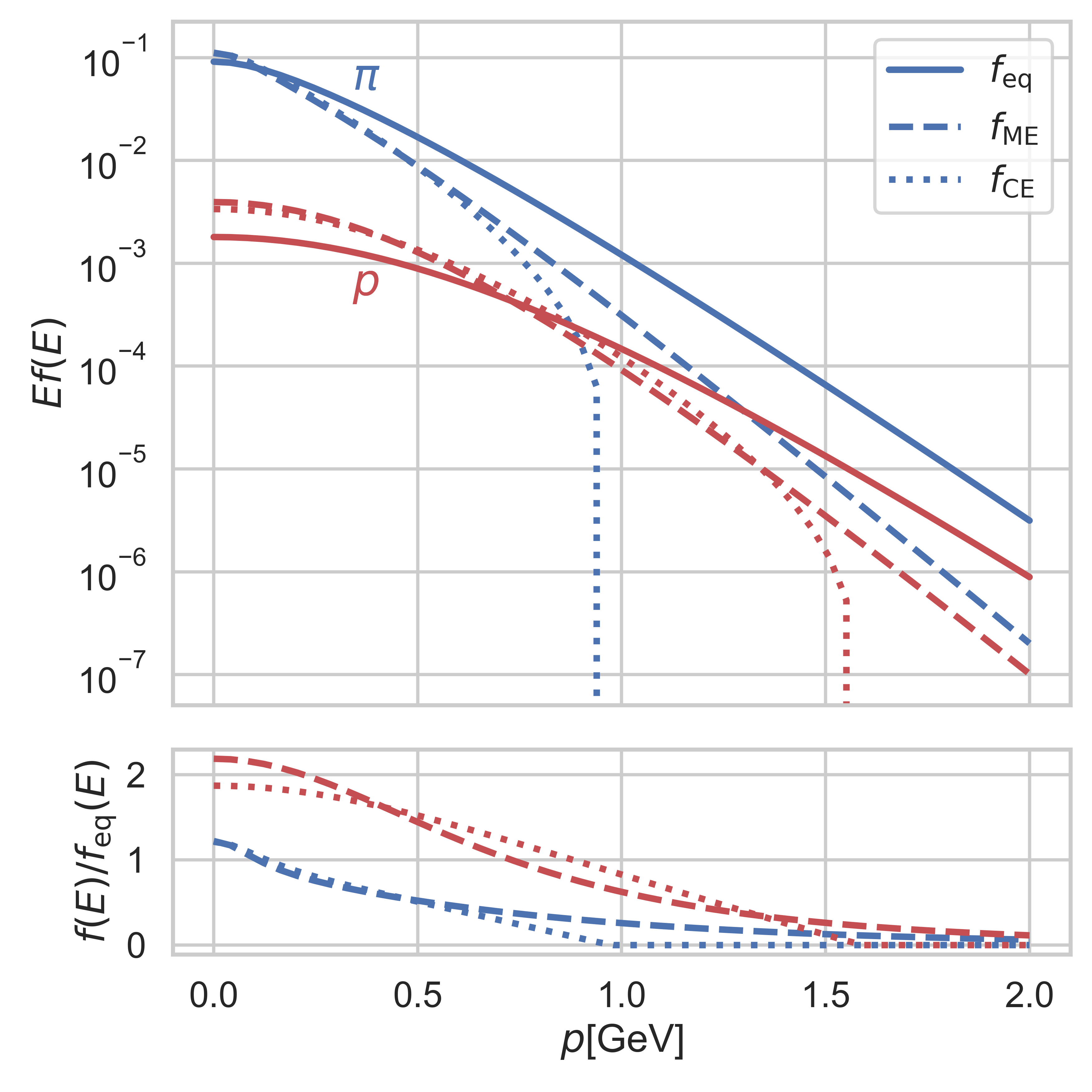}\\[-3ex]
\caption{
    The particle spectra in the local fluid rest frame for the local equilibrium (solid), maximum-entropy (ME, dashed) and linear Chapman-Enskog RTA (CE, dotted) distributions, for pions (blue) and protons (red). The bottom panel displays the ratio of these spectra to the local equilibrium spectra.
    \vspace*{-3mm}}
\label{fig:lrf_spectra}
\end{figure}

We now compare the maximum-entropy distribution with the linear Chapman-Enskog RTA distribution and the local-equilibrium distribution, all in the local rest frame. The linear Chapman-Enskog RTA viscous correction is given by \cite{Bozek:2009dw, McNelis:2019auj, Jaiswal:2014isa, Dusling:2011fd}
\begin{equation}
    \delta f_{\rm CE} = f_{\rm eq} \bar{f}_{\rm eq}
    \left[  \frac{\Pi}{\beta_{\Pi}}\left( \frac{(u{\,\cdot\,}p) \mathcal{F}}{T^2}
    - \frac{p{\,\cdot\,}\Delta{\,\cdot\,}p}
    {3(u{\,\cdot\,}p)T}\right) \right],
\end{equation}
where the coefficients $\mathcal{F}$ and $\beta_{\Pi}$ are given by thermodynamic integrals of the hadron gas:
\begin{equation}
    \mathcal{F} \equiv -T^2 \frac{\epsilon + p_{\rm eq}}{\mathcal{J}_{(3, 0)}},\quad
    \beta_{\Pi} \equiv \mathcal{F} \frac{\epsilon + p_{\rm eq}}{T} + \frac{5 J_{(3, 2)}}{3T},
\end{equation}
where the moments $J_{(n,q)}$ are defined by
\begin{equation}
    J_{(n,q)} \equiv \frac{1}{(2q + 1)!!} \int_p (u \cdot p)^{n-2q} \, \bigl(p^{\langle\alpha\rangle} p_{\langle\alpha\rangle}\bigr)^{q} \, \feqh \feqhbar.
\end{equation}
For our comparison we consider a hadron resonance gas at temperature $T = 0.15$\,GeV. We assume a moderately large, negative bulk pressure with magnitude of one-third of the hadron resonance gas equilibrium pressure at this temperature: $\Pi = -p_{\rm eq} / 3$. For a bulk pressure of this magnitude the Chapman-Enskog RTA linear bulk correction becomes larger than the ideal part, $|\delta f_{h, \rm CE}| \gtrsim f_{h, \rm eq}$, already at moderate values of momentum $|\bm{p}| \sim 1$ GeV. In practice, when doing particlization in simulations of heavy ion collisions, the viscous correction must be regulated by hand. This is required to maintain positivity of the distribution function, which is interpreted as a probability distribution from which particles and their momenta are sampled. A typical procedure is to replace 
\begin{equation}
    \delta f_{\rm CE} \rightarrow \sign(\delta f_{\rm CE}) \min( f_{\rm eq}, |\delta f_{\rm CE}| ) \equiv \delta \tilde{f}_{\rm CE},
\end{equation}
which we will call the `regulated Chapman-Enskog viscous correction'. We note that, in practice, this regulation breaks the exact matching of the dissipative part of the stress-tensor:
\begin{equation}
    \delta T^{\mu\nu} = \int_p p^{\mu} p^{\nu} \delta f_{h, \rm CE} \neq \int_p p^{\mu} p^{\nu} \delta \tilde{f}_{h, \rm CE}.
\end{equation}
Restoring the exact matching condition would require recalculating the coefficients $\mathcal{F}$ and $\beta_{\Pi}$ using the regulated Chapman-Enskog RTA correction form in the integrands. 

In Fig.~\ref{fig:lrf_spectra} we compare the maximum-entropy distribution with the regulated CE RTA distribution and the local equilibrium distribution. The deviations from the equilibrium distribution are large. For pions, the viscous corrections are negative, except at very low momenta $p<50$\,MeV. The CE RTA and ME distributions agree well at low momenta but disagree dramatically at $p>1$\,GeV where the (unregulated) CE RTA distribution goes negative. For the heavier protons, the bulk viscous corrections are much larger and switch sign at intermediate momenta ($p\sim 0.7{-}0.9$\,GeV), being positive at lower and negative at larger momenta. Significant discrepancies between the two viscous distributions are observed for protons over the entire momentum range; at $p=0$, the ME proton distribution is about 20\% larger than the CE RTA distribution and more than a factor 2 larger than the equilibrium distribution.

\begin{figure}
\includegraphics[width=8cm]{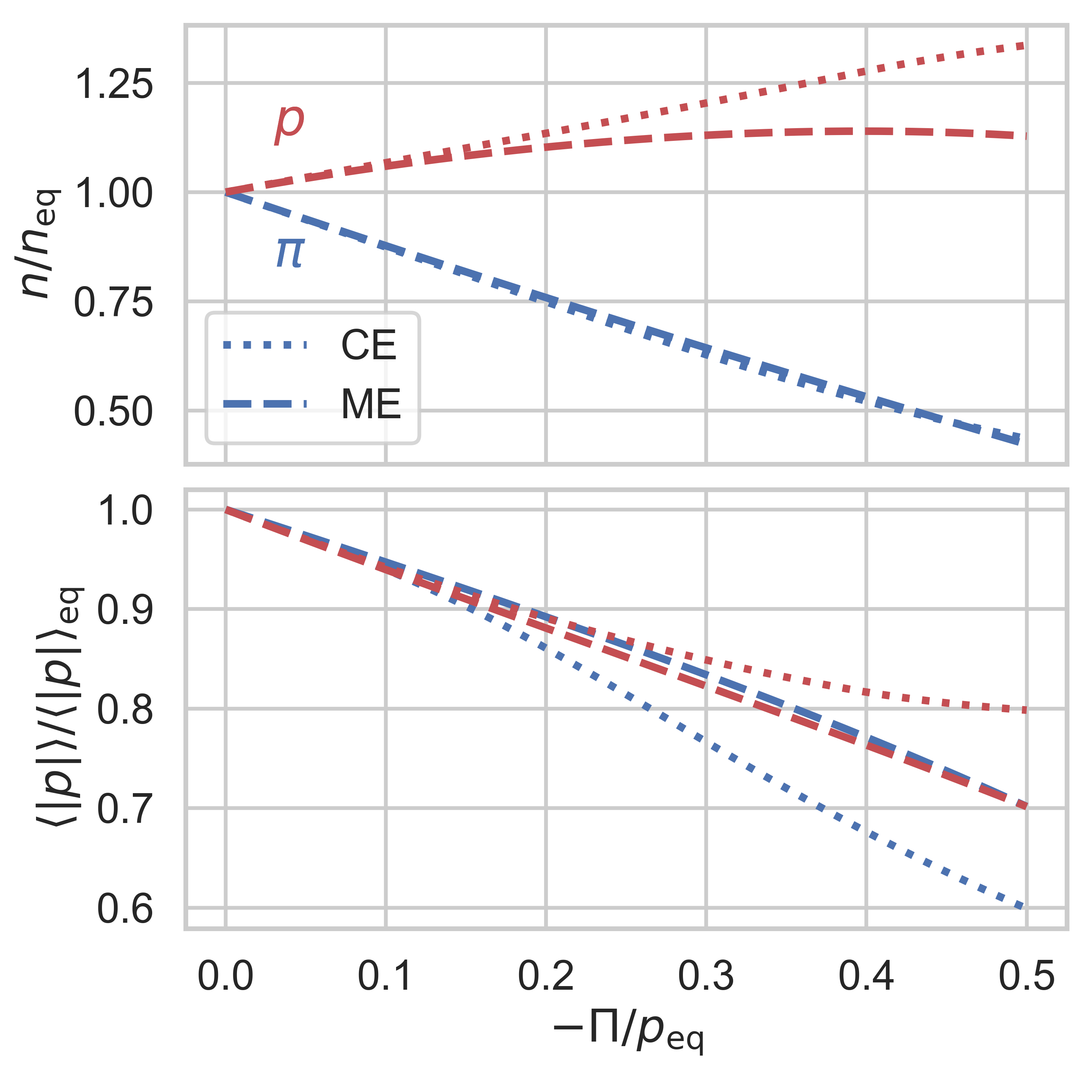}\\[-3ex]
\caption{
    {\sl Top:} Particle densities in the local rest frame, normalized by their equilibrium values, for the maximum-entropy distribution (ME, dashed) and linear Chapman-Enskog RTA distribution (CE, dotted), for pions (blue, decreasing) and protons (red, increasing), as functions of the normalized bulk viscous pressure $-\Pi/p_\mathrm{eq}$. {\sl Bottom:} The corresponding mean particle momenta in the local rest frame, normalized by their equilibrium values.
    \vspace*{-5mm}}
\label{fig:lrf_densities}
\end{figure}

In Fig.~\ref{fig:lrf_densities} we plot the local rest frame particle densities and mean momentum magnitudes, both normalized by their equilibrium values, for pions and protons as functions of the bulk inverse Reynolds number $-\Pi / p_{\rm eq}$. We see that for protons the maximum-entropy distribution yields smaller bulk viscous corrections to the particle density than the regulated linear Chapman-Enskog RTA result. For pions the bulk viscous corrections to the particle densities have the opposite sign and are much larger, but the predictions of the ME and regulated CE RTA distributions agree well with each other, even for large values of the bulk viscous pressure. The same is not true for the pion and proton mean momenta which, for large bulk inverse Reynolds numbers, differ significantly, in opposite directions, between the maximum-entropy and regulated Chapman-Enskog viscous distributions. The differences in the proton yields and pion and proton mean momenta between these two different ans\"atze for the bulk viscous distribution functions are large enough to have the potential of significantly affecting the bulk viscosity inferred from model-to-data comparisons. 

As a final note we observe that the maximum-entropy distribution can naturally handle very large bulk inverse Reynolds numbers. Physically, the bulk inverse Reynolds number may grow large near the pseudo-critical temperature $T_c \sim 150$\,MeV where the quark-gluon plasma turns into hadrons \cite{Karsch:2007jc, NoronhaHostler:2008ju, Arnold:2006fz}. Since the maximum-entropy method is not based on a near-equilibrium expansion, it does not require the dissipative stresses to be small for self-consistency. 

\vspace*{-3mm}
\section{Matching with shear stress}
\label{sec5}
\vspace*{-3mm}

\subsection{The relationship between the shear stress and its Lagrange multipliers}
\label{sec5a}
\vspace*{-3mm}

In this section, we again consider the simpler case of Maxwell-Boltzmann particles ($\theta = 0$). We return to the matching condition (\ref{eqn:shear_match}) for the shear stress tensor and write it in the local rest frame: 
\begin{equation}
\label{match_pi}
    \pi^{ij} = \Delta^{ij}_{kl}
    \int_p p^{k}p^{l} \exp\Bigl(-\Lambda p_0 - \frac{\lambdapi}{p_0} \bm{p}^2
   \Bigr) \exp \Bigl( - \frac{\gamma_{rs} p^r p^s}{p_0} \Bigr).
\end{equation}
Latin tensor indices run over the spatial directions in the LRF. 

The matching condition (\ref{match_pi}) establishes a highly nonlinear relationship between the given shear stress in the LRF, $\pi^{ij}$, and the associated symmetric and traceless 3$\times$3 matrix of Lagrange multipliers, $\gamma^{ij}$. We now proceed to show that these two matrices share a common set of eigenvectors. This will be seen to decisively simplify the task of determining the Lagrange multipliers $\gamma^{ij}$ from the shear stress $\pi^{ij}$.

Let us Taylor expand the second exponential in (\ref{match_pi}),
\begin{equation} 
\label{eqn:expandexp}
   \exp \Bigl( - \frac{\gamma_{rs} p^r p^s}{p_0} \Bigr) = \sum_{n=0}^\infty \Bigl(- \frac{ \gamma_{ij}p^{i}p^{j}}{p_0} \Bigr)^n,
\end{equation}
and consider truncating this series at some finite order. Truncating at $n=0$ yields $\pi^{ij}_{(0)}=0.$ At truncation order $n=1$ we get
\begin{eqnarray}
    \pi^{ij}_{(1)}&=& -\Delta^{ij}_{kl} \gamma_{ab} \int_p \frac{1}{p_0} \, p^k \, p^l \, p^a \, p^b \, \exp\Bigl(-\Lambda p_0 - \frac{\lambdapi}{p_0} \bm{p}^2
   \Bigr) 
\nonumber\\
   &=& -\Delta^{ij}_{kl} C_{1,0} (\Lambda, \lambdapi) \, \gamma^{kl},
\end{eqnarray}
where $C_{1,0}(\Lambda, \lambdapi)$ is a scalar function. In $C_{r,s}$, the first index $r$ denotes the total number of tensors $\bm{\gamma}$ following the coefficient while the second index $s$ denotes how many of these $\bm{\gamma}$ tensors have their indices mutually contracted to form scalars. (This will become clearer below.)
At $n = 2$ we encounter the integral
\begin{eqnarray}
    &&\Delta^{ij}_{kl} \gamma_{ab}\, \gamma_{cd} \int_p \frac{1}{p_0^2} \, p^k \, p^l \, p^a \, p^b \, p^c \, p^d \exp\Bigl(-\Lambda p_0 - \frac{\lambdapi}{p_0} \bm{p}^2 \Bigr) 
\nonumber\\
    &&= \Delta^{ij}_{kl}\, C_{2,0} (\Lambda, \lambdapi) \, (\bm{\gamma}^2)^{kl},
\end{eqnarray}
where $(\bm{\gamma}^2)^{kl}=\gamma^k_c\gamma^{cl}$. (Note that the term $C_{2,1} \gamma^{kl} \, \gamma^{a}_{a}$ vanishes because $\bm{\gamma}$ is traceless.) The integral over the $n = 3$ term in the series yields
\begin{equation}
  -\Delta^{ij}_{kl} \, \Bigl( C_{3,0} \, (\bm{\gamma}^3)^{kl} + C_{3,2} \, \gamma^{kl} \, \gamma^2 \Bigr),
\end{equation}
where we introduced $\gamma^n\equiv \mathrm{tr}(\bm{\gamma}^n)$, and the $n = 4$ term similarly integrates to
\begin{equation}
    \Delta^{ij}_{kl} \bigl( C_{4,0} \, (\bm{\gamma}^4)^{kl}
     + C_{4,2} \, (\bm{\gamma}^2)^{kl}\,\gamma^2 + C_{4,3}\,\gamma^{kl} \,\gamma^3 \bigr).
\end{equation}
It is clear that the $n^{\rm th}$ order introduces one new term
\begin{equation}
     (-1)^n\,\Delta^{ij}_{kl} \, C_{(n,0)} (\bm{\gamma}^n)^{kl}
\end{equation}
while all other terms have the same tensor structure as lower order terms. For example, when we truncate Eq.~(\ref{eqn:expandexp}) at order $n = 4$ we obtain
\begin{eqnarray}
    \pi^{ij}_{(4)} &=& \Delta^{ij}_{kl}\, \bigl(- c_1^{(4)}  \bm{\gamma} + c_2^{(4)} \bm{\gamma}^2 
    - c_3^{(4)} \bm{\gamma}^3 + c_4^{(4)} \bm{\gamma}^4 \bigr)^{kl}
\end{eqnarray}\\
with the coefficients
\begin{align*}
    c_1^{(4)} &= C_{1,0} + C_{3,2} \gamma^2 + C_{4,3} \gamma^3 ,
\\
    c_2^{(4)} &= C_{2,0} + C_{4,2} \gamma^2,\quad
    c_3^{(4)} = C_{3,0},\quad  
    c_4^{(4)} = C_{4,0}.
\end{align*}
The scalar coefficients $c_i$ are functions of $\Lambda$, $\lambdapi$, and scalar contractions of $\bm{\gamma}$.

Summed to all orders the series for $\pi^{ij}$ thus becomes
\begin{equation} 
\label{seriesingamma}
    \pi^{ij} = \Delta^{ij}_{kl}\, \Bigl(\sum_{n=1}^\infty
    \tilde{c}_n \, (-\bm{\gamma})^n\Bigr)^{kl},
\end{equation}
where the coefficients $\tilde{c}_n$ include contributions $c_n^{(i)}$ from all orders $1\leq i\leq \infty$. Note that although $\gamma^{ij}$ is traceless, products of $\gamma^{ij}$'s are not. The operator $\Delta^{ij}_{kl}$ projects out the trace part of such products. For example,
\begin{eqnarray}
    \Delta^{ij}_{kl} \, (\bm{\gamma}^2)^{kl} &=& 
    \gamma^{i}_{a} \, \gamma^{a \, j} - \frac{\delta^{ij}}{3} \gamma^{a}_{b} \, \gamma^{b}_{a}
    = (\bm{\gamma}^2)^{ij}
    - \frac{\delta^{ij}}{3}\gamma^2.\quad
\end{eqnarray}

Equation~(\ref{seriesingamma}) can be written in matrix notation as
\begin{equation}
    \bm{\pi} = \bm{\Gamma} - {\textstyle\frac{1}{3}} \mathrm{tr} (\bm{\Gamma})\,\bm{I},
\end{equation}
where $\bm{I}$ is the $3\times 3$ identity matrix and
\begin{equation}
   \bm{\Gamma} \equiv \sum_{n=1}^\infty (-1)^n
    \tilde{c}_n \, \bm{\gamma}^n. 
\end{equation}
Since any power of a matrix commutes with itself, $[\bm{\gamma}, \bm{\gamma}^n] = 0$, it follows that $[\bm{\gamma}, \bm{\pi}] = 0$. Therefore, the shear stress tensor $\bm{\pi}$ and its associated tensor of Lagrange multipliers $\bm{\gamma}$ are simultaneously diagonalisable.

At any spacetime point $x$, the hydrodynamic energy-momentum tensor $T^{\mu\nu}(x)$ provides us (after transformation to the LRF at $x$) with the full matrix $\bm{\pi}$. Finding the eigenvalues and eigenvectors of the given $\bm{\pi}$ is straightforward. As a real, symmetric and traceless matrix, $\bm{\pi}$ has two independent real eigenvalues, associated with three  orthonormal eigenvectors. [It is easy to show that orthogonality and normalization allow to characterize the three eigenvectors by three real parameters (Euler angles). Together with the two independent eigenvalues we thus recover the five independent shear stress degrees of freedom.] The matrix $\bm{C}$ of orthonormal eigenvectors can be used to rotate $\bm{\pi}$ into diagonal form:
\begin{equation}
\label{eqn:pi_eig}
    \bm{\pi} = \bm{C^T \pi_{_\mathrm{D}} C}.
\end{equation}
Here $\bm{\pi_{_\mathrm{D}}} \equiv \mathrm{diag} (\pi_1, \pi_2, {-}\pi_1{-}\pi_2)$ is the diagonalized shear stress, with eigenvalues $\pi_1$, $\pi_2$, and $\pi_3=-(\pi_1{+}\pi_2)$. From the arguments above it follows that $\bm{\gamma}$ is diagonalized by the same transformation:
\begin{equation}
\label{eqn:gamma_eig}
    \bm{\gamma} = \bm{C^T \gamma_{_\mathrm{D}} C}.
\end{equation}
This implies that three of the five independent Lagrange multipliers can be determined easily from the eigenvectors of the shear stress. Since the rotation matrix $\bm{C}$ is known from the diagonalization (\ref{eqn:pi_eig}) of the given shear stress $\bm{\pi}$, the full Lagrange multiplier matrix $\bm{\gamma}$ is easily computed from (\ref{eqn:gamma_eig}) once the two independent diagonal elements of $\bm{\gamma}_{_\mathrm{D}}$ have been determined. The only non-linear part of the problem is the computation of the two independent eigenvalues of $\bm{\gamma}$ from those of $\bm{\pi}$. These statements hold irrespective of the value of the bulk viscous pressure. They greatly simplify the numerical task of finding the Lagrange multiplier matrix $\bm{\gamma}$ that matches the given shear stress $\bm{\pi}$.\footnote{%
    The idea to simplify the solution of the matching conditions by computing the hydrodynamic moments of the anisotropic distribution function in a frame which diagonalizes the shear tensor was also exploited in Ref.~\cite{Nopoush:2019vqc}.}  

\vspace*{-3mm}
\subsection{Solution for a massless gas}
\vspace*{-2mm}

In this Section we illustrate the determination of the Lagrange multipliers $\gamma^{ij}$ in the maximum-entropy distribution (\ref{f_h},\ref{f_h_LRF}) from $\pi^{ij}$ for the case of a massless Maxwell-Boltzmann gas with nonzero shear stress but vanishing bulk viscous pressure. By keeping the discussion initially general we show that the future generalization of this solution to a general gas mixture of massive hadronic resonances characterized by non-zero values for both shear {\it and} bulk viscous stresses will be straightforward.

Let us return to the generating function $\mathcal{Z}$ from Eq.~(\ref{genfunc0}), now expressed covariantly and  specifically for Maxwell-Boltzmann particles:
\begin{widetext}
\begin{equation}
\label{genfunc}
    \mathcal{Z}(\Lambda, \lambdapi, \gamma_{\alpha \beta}) \equiv \int_p (u \cdot p) \exp\bigl[-\Lambda (u \cdot p)\bigr] \,\exp\Bigl[\lambdapi \frac{p\cdot\Delta\cdot p}{u\cdot p}\Bigr] \,\exp \left[-\gamma_{\alpha \beta} \frac{p^{\langle \alpha}p^{\beta \rangle}}{u \cdot p} \right].
\end{equation}
\end{widetext}
According to Eq.~(\ref{derivatives}), the energy density, total isotropic pressure and shear stress from Eqs.~(\ref{e})--(\ref{eqn:shear_match}) are then given by the following derivatives with respect to the Lagrange multipliers:
\begin{equation}
\label{derivatives1}
    \epsilon = - \frac{\partial \mathcal{Z}}{\partial \Lambda},\quad
    P = -\frac{1}{3}\frac{\partial \mathcal{Z}}{\partial \lambdapi},\quad
    \pi^{\mu\nu} = - \frac{\partial \mathcal{Z}}{\partial \gamma_{\mu\nu}}.   
\end{equation}
Note that in the generating functional (\ref{genfunc}) we do not impose tracelessness and transversality on $\gamma_{\alpha\beta}$, i.e. when taking the derivatives (\ref{derivatives1}) we consider, in particular, all three eigenvalues $\gamma_1$, $\gamma_2$, and $\gamma_3$ as independent. In the last equation (\ref{derivatives1}), the correct symmetries of $\pi^{\mu\nu}$ are ensured by the spatial and transverse projection implied by the angular parentheses in the factor $p^{\langle \alpha}p^{\beta \rangle}$ in Eq.~(\ref{genfunc}).

As before, we will work out these expressions in the local rest frame. However, to simplify the last factor in (\ref{genfunc}) we will rotate the LRF integration momentum variables with the matrix $\bm{C}$ that diagonalizes $\bm{\gamma}$. The integration measure $\int_p$ and the first three factors under the integral (\ref{genfunc}) are invariant under this rotation. Writing $\vec{q}=\bm{C}\vec{p}\,$ as well as $q\equiv|\vec{q}|$, $q_0=\sqrt{q^2{+}m^2}$, and computing
\begin{equation}
    \gamma_{\langle ij \rangle}p^{\langle i}p^{j\rangle} = \gamma_1 q_1^2 + \gamma_2 q_2^2 + \gamma_3 q_3^2 - \frac{q^2}{3} (\gamma_1 + \gamma_2 + \gamma_3)
\end{equation}
(the tracelessness condition $\gamma_3=-(\gamma_1{+}\gamma_2)$ will only be implemented at the end), the generating function (\ref{genfunc}) takes the form
\begin{eqnarray}
\label{genfunc2}
    &&\mathcal{Z} = \sum_h \frac{g_h}{(2\pi)^3} \int_0^\infty\!\!\! q^2 dq \,e^{-\Lambda q_0}\, 
    {\exp}\left[\frac{q^2}{q_0}\left(\frac{\gamma_1{+}\gamma_2{+}\gamma_3}{3}{-}\lambdapi\right)\right]
\nonumber\\
    &&\times
    \int_{\Omega_q} \!\!\!\!
    {\exp}\left[-\frac{q^2}{q_0}\Bigl (\sin^2\theta_q\bigl({\cos^2}\phi_q \gamma_1 {+} {\sin^2}\phi_q\gamma_2\bigr) + {\cos^2}\theta_q\gamma_3\Bigr) \right].
\nonumber\\
\end{eqnarray}
Here $(\theta_q,\phi_q)$ are the polar and azimuthal angles of $\vec{q}$, with standard integration measure $d\Omega_q=\sin\theta_q d\theta_q d\phi_q$. From Eq.~(\ref{genfunc2}) the generating function $\mathcal{Z}$ can be calculated numerically. Taking its derivatives (\ref{derivatives1}) with respect to the four independent Lagrange multipliers on which it depends and solving the resulting matching conditions iteratively may be a difficult task whose full solution will be left to future work. In the following two subsections we work out the generating function semi-analytically for the simpler case of a massless Boltzmann gas (where $m{=}0$ and $q_0=q=|\vec{q}|$) and then evaluate the matching conditions in this simplified setting.

\vspace*{-4mm}
\subsubsection{Evaluation of $\mathcal{Z}$ for a massless Boltzmann gas}
\vspace*{-3mm}

For a single-species Boltzmann gas of spinless and massless particles the generating function (\ref{genfunc2}) reduces to
\begin{eqnarray}
    {\cal Z} &=& \int_{0}^{\infty} \frac{q^2 dq}{(2\pi)^3} \, e^{-\bar{\Lambda} q} \, \int_{-1}^{1} dz\, e^{- q \gamma_3 z^2} 
\\\nonumber
    &\times& \int_{0}^{2\pi} d\phi_q \, e^{-q (1{-}z^2) \bigl({\cos^2}\phi_q \gamma_1 {+} {\sin^2}\phi_q  \gamma_2\bigr)},
\end{eqnarray}
\vspace*{2mm}
with $z{\,=\,}\cos\theta_q$ and $\bar\Lambda{\,=\,}\Lambda{+}\lambdapi - \frac{1}{3}(\gamma_1 {+} \gamma_2 {+} \gamma_3)$. The azimuthal integral yields a modified Bessel function:
\begin{equation}
\nonumber
    \int_{0}^{2\pi}\!\!\!d\phi_q (\dots) = 2\pi e^{-q(1{-}z^2)(\gamma_1{+}\gamma_2)/2} \, I_0\Bigl(q(1{-}z^2)\frac{\gamma_1{-}\gamma_2}{2}\Bigr). 
\end{equation}
%
%
\begin{equation}
    \!\!\!\!\!\text{Hence}\qquad
    {\cal Z} = 
    \int_{0}^{\infty} \frac{q^2 dq}{(2\pi)^2} \, \int_{-1}^{1} dz\,e^{-q\Lambda'(z)} I_0\bigl(q\alpha(z)\bigr),
 \end{equation}
where $\Lambda'(z) \equiv \bar{\Lambda} + \frac{1}{2}(1{-}z^2)(\gamma_1{+}\gamma_2)+ 
z^2 \gamma_3$ and $\alpha(z) \equiv \frac{1}{2}(1{-}z^2)(\gamma_1{-}\gamma_2)$. The $q$ integral is done using
\begin{equation}
    \int_{0}^{\infty} dx \, x^2 \, e^{-\Lambda' x} \, I_0(\alpha x) = 
    \frac{2 (\Lambda')^2 + \alpha^2}{\bigl[(\Lambda')^2 - \alpha^2\bigr]^{5/2}},
\end{equation}
leaving us with the following 1-dimensional polar-angle integral for the generating function: 
\begin{equation}
\label{Zmethod1}
    \!\!\!
    {\cal Z}(\Lambda,\lambdapi,\gamma_1,\gamma_2,\gamma_3) =\! \int_{0}^{1} \! \frac{dz}{2\pi^2} \, 
    \frac{2 (\Lambda'(z))^2 + \alpha^2(z)}{\bigl[(\Lambda'(z))^2 {-} \alpha^2(z)\bigr]^{5/2}}.
\end{equation}
Its derivatives with respect to $\Lambda$, $\lambdapi$, $\gamma_1$ and $\gamma_2$ yield a coupled set of four matching conditions, with the right hand sides given by 1-dimensional integrals of the structure (\ref{Zmethod1}).\footnote{%
    This is the point where we set $\gamma_3=-(\gamma_1{+}\gamma_2)$.}
This is certainly easier than working from Eq.~(\ref{genfunc2}) which involves 3-d integrals over more complex integrands. We will explore in the next subsection an alternate approach based on series expansions.

\vspace*{-4mm}
\subsubsection{Alternate method}
\vspace*{-3mm}

The following approach uses a series expansion that allows to perform the angular integrals even in the general case of a gas mixture of massive hadrons. Writing the exponential under the angular integral in Eq.~(\ref{genfunc2}) as
\begin{eqnarray}
\label{expansion}
    &&e^{-(q_1^2\gamma_1{+}q_2^2\gamma_2{+}q_3^2\gamma_3)/q_0} = 
\\\nonumber
    &&
    \sum_{n_1,n_2,n_3}\frac{1}{n_1!}  \Bigl(\frac{-\gamma_1 q_1^2}{q_0}\Bigr)^{\!n_1} \frac{1}{n_2!}  \Bigl(\frac{-\gamma_2 q_2^2}{q_0}\Bigr)^{\!n_2}  \frac{1}{n_3!}  \Bigl(\frac{-\gamma_3 q_3^2}{q_0}\Bigr)^{\!n_3} ,
\end{eqnarray}
where $q_1=q\sin\theta_q\cos\phi_q$, $q_2=q\sin\theta_q\sin\phi_q$, and $q_3=q\cos\theta_q$, the angular integrals over each term can be done:
\begin{eqnarray}
    &&\int d\Omega\, (\cos\phi)^{2n_1} (\sin\phi)^{2n_2} (\cos\theta)^{2n_3} (\sin\theta)^{2(n_1+n_2)} \equiv 
\nonumber \\
    && \mathcal{A}(n_1, n_2, n_3) 
    = 2\, \frac{ \Gamma(n_1+\frac{1}{2}) \Gamma(n_2+\frac{1}{2}) \Gamma(n_3+\frac{1}{2})}
               {\Gamma(n_s+\frac{3}{2})},\quad
\end{eqnarray}
where we introduced $n_s \equiv n_1{+}n_2{+}n_3$ for brevity. This yields the following form for the generating function:
\begin{widetext}
\begin{equation}
    \!\!\!\!
    \mathcal{Z} = \sum_h\frac{g_h}{(2\pi)^3} \int q^2 dq\, e^{-\Lambda q_0 -\bigl(\lambdapi -\frac{1}{3}(\gamma_1{+}\gamma_2{+}\gamma_3)\bigr) q^2/q_0} \sum_{n_1, n_2, n_3} \frac{\mathcal{A}(n_1, n_2, n_3)}{n_1!\,n_2!\,n_3!}  (-\gamma_1)^{n_1} (-\gamma_2)^{n_2} (-\gamma_3)^{n_3} \left(\frac{q^2}{q_0}\right)^{\!n_s}.
\end{equation}
This expression is still valid for a massive gas mixture with classical Boltzmann statistics and can thus form the basis for future generalizations of what we derive below. Please note that we have not yet used the zero-trace condition $\gamma_1{+}\gamma_2{+}\gamma_3=0$; as before, we only implement it at the end.

We now restrict our attention again to the simpler case of massless particles ($q_0=q$) for which the momentum integral is easily performed:
\begin{equation}\label{Z_series}
    \mathcal{Z} = \frac{1}{(2\pi \bar{\Lambda} )^3} \sum_{n_1, n_2, n_3} \frac{\mathcal{A}(n_1, n_2, n_3)}{n_1!\,n_2!\,n_3!} 
    (n_s{+}2)! (-1)^{n_s} \left(\frac{\gamma_1}{\bar{\Lambda}}\right)^{n_1}  \left(\frac{\gamma_2}{\bar{\Lambda}}\right)^{n_2} \left(\frac{\gamma_3}{\bar{\Lambda}}\right)^{n_3}  .
\end{equation}
Here $\bar\Lambda=\Lambda+\lambdapi-\frac{1}{3}(\gamma_1{+}\gamma_2{+}\gamma_3)$. By keeping a sufficient number of terms in the series given above, we have checked that 
both Eq. (\ref{Z_series}) and Eq. (\ref{Zmethod1}) yield identical results for ${\cal Z}$ for a given set of Lagrange parameters. The derivative with respect to the eigenvalue $\gamma_1$ ($\gamma_2$) of the Lagrange multipler tensor $\bm{\gamma}$ yields the eigenvalue $\pi_1$ ($\pi_2$) of the shear stress tensor:
\begin{eqnarray}
\label{eqn:pi_1_series}
    \pi_1 &=& -\frac{\partial \mathcal{Z}}{\partial \gamma_1} =  - \frac{1}{(2\pi)^3 \bar{\Lambda}^4} \sum_{n_1, n_2, n_3} 
    \frac{\mathcal{A}(n_1, n_2, n_3)}{n_1!\,n_2!\,n_3!} 
    (n_s{+}2)!
    (-1)^{n_s}  \left(\frac{\gamma_1}{\bar{\Lambda}}\right)^{n_1}  \left(\frac{\gamma_2}{\bar{\Lambda}}\right)^{n_2} \left(\frac{\gamma_3}{\bar{\Lambda}}\right)^{n_3} \left( 1 + \frac{n_s}{3} + n_1 \frac{\bar{\Lambda}}{\gamma_1} \right),
\\
\label{eqn:pi_2_series}
    \pi_2 &=& -\frac{\partial \mathcal{Z}}{\partial \gamma_2} =   - \frac{1}{(2\pi)^3 \bar{\Lambda}^4} \sum_{n_1, n_2, n_3} \frac{\mathcal{A}(n_1, n_2, n_3)}{n_1!\,n_2!\,n_3!} 
    (n_s{+}2)! (-1)^{n_s} \left(\frac{\gamma_1}{\bar{\Lambda}}\right)^{n_1}  \left(\frac{\gamma_2}{\bar{\Lambda}}\right)^{n_2} \left(\frac{\gamma_3}{\bar{\Lambda}}\right)^{n_3} \left( 1 + \frac{n_s}{3} + n_2 \frac{\bar{\Lambda}}{\gamma_2} \right).
\end{eqnarray}
The energy density and total isotropic pressure are given by
\begin{eqnarray}
\label{eqn:eps_series}
    \epsilon &=& -\frac{\partial \mathcal{Z}}{\partial \Lambda} = \frac{1}{(2\pi)^3 \bar{\Lambda}^4} \sum_{n_1, n_2, n_3} 
    \frac{\mathcal{A}(n_1, n_2, n_3)}{n_1!\,n_2!\,n_3!} 
    (n_s{+}3)! (-1)^{n_s}
\left(\frac{\gamma_1}{\bar{\Lambda}}\right)^{n_1}  \left(\frac{\gamma_2}{\bar{\Lambda}}\right)^{n_2} \left(\frac{\gamma_3}{\bar{\Lambda}}\right)^{n_3} , 
\\
\label{eqn:Pseries}
    P &=& p_\mathrm{eq}(\epsilon) + \Pi 
    =
    -\frac{1}{3}\frac{\partial \mathcal{Z}}{\partial \lambdapi} = \frac{1}{3(2\pi)^3 \bar{\Lambda}^4} \sum_{n_1, n_2, n_3} \frac{\mathcal{A}(n_1, n_2, n_3)}{n_1!\,n_2!\,n_3!} (n_s{+}3)! (-1)^{n_s} \left(\frac{\gamma_1}{\bar{\Lambda}}\right)^{n_1}  \left(\frac{\gamma_2}{\bar{\Lambda}}\right)^{n_2} \left(\frac{\gamma_3}{\bar{\Lambda}}\right)^{n_3}. 
\end{eqnarray}
\end{widetext}
All of these expressions are to be evaluated at $\gamma_3=-(\gamma_1{+}\gamma_2)$ and $\bar\Lambda=\Lambda+\lambdapi$. The series in the last two equations are seen to be related by $P=\epsilon/3$ which (with the equation of state for a massless gas $p_\mathrm{eq}(\epsilon)=\epsilon/3$) yields $\Pi=0$, as it should: in a massless gas the bulk viscous pressure vanishes. Correspondingly, $\lambdapi=0$ for this system.

\vspace*{-3mm}
\subsection{Results for a massless Boltzmann gas}
\vspace*{-2mm}

This leaves us with the coupled equations (\ref{eqn:pi_1_series})--(\ref{eqn:eps_series}), evaluated at $\bar\Lambda=\Lambda$ and $\gamma_3=-(\gamma_1{+}\gamma_2)$. They can be inverted
numerically with a three-dimensional root solving method\footnote{%
    The code used in this section can be found in \cite{ME_git}.} 
to find $(\Lambda, \gamma_1, \gamma_2)$ from $(\epsilon,\pi_1,\pi_2)$. Initializing the root solver with a guess given by the linearized CE RTA expressions (\ref{eqn:CE_RTA_coeffs}) was found to reduce the number of iterations needed. 

The algorithm and root solver were tested in a blind test where one of the authors selected a value of $\Lambda$ and a matrix $\bm{\gamma}$ of shear stress Lagrange multipliers in the LRF, without any symmetry restrictions, used Eqs.~(\ref{eqn:ematch}) and (\ref{pimatch}) (with $\lambdapi=0$) to generate the corresponding energy density $\epsilon$ and the full LRF shear stress tensor $\pi^{ij}$ for a massless Boltzmann gas, and handed these to another author who then diagonalized $\pi^{ij}$, used Eqs.~(\ref{eqn:pi_1_series})--(\ref{eqn:eps_series}) to 
reconstruct $(\Lambda, \gamma_1, \gamma_2)$ and, finally, the complete Lagrange multiplier tensor $\bm{\gamma}$ from Eq.~(\ref{eqn:gamma_eig}). The reconstructed Lagrange multipliers agreed with the originally selected ones to a precision that can be systematically improved by truncating the numerical series at higher orders.\footnote{%
    For an inverse shear Reynolds number of 0.2 we found per mille agreement by truncating at $n_i=10$ ($i=1,2,3$).}
%
%
\begin{figure}[t!]
\includegraphics[width=0.98\linewidth]{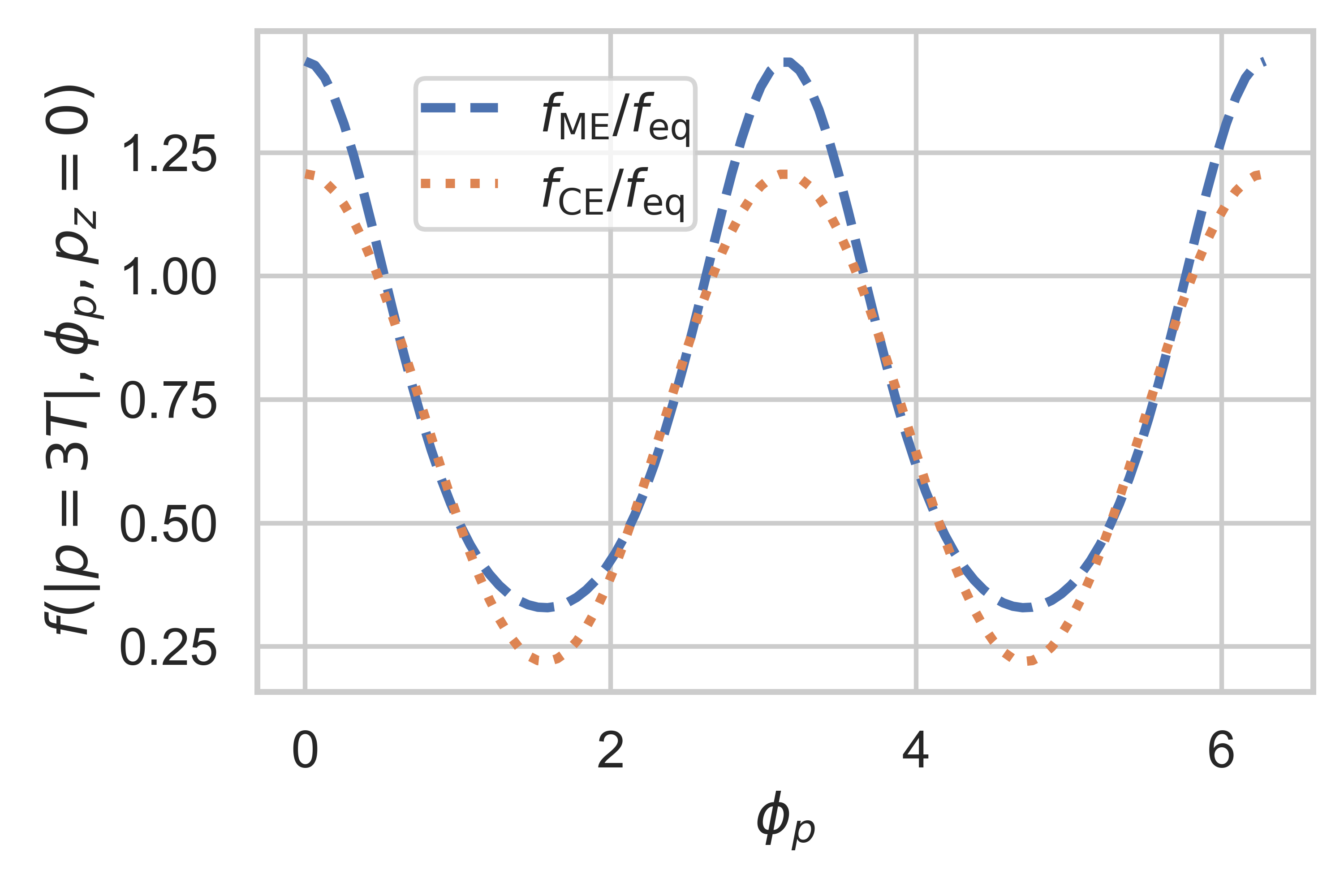}\\[-3ex]
\caption{
    The maximum-entropy (ME, blue dashed) and linear Chapman-Enskog RTA (CE, red dotted) distributions in the LRF as functions of the azimuthal momentum angle $\phi_p$ for a diagonal shear stress with $(x, z)$ isotropy, taking $\pi^{xx} = \pi^{zz} = p_{\rm eq} / 5$. Both are normalized by the equilibrium distribution.
    \vspace*{-3mm}}
\label{fig:shear_phip}
\end{figure}
%
\begin{figure}[!hb]
\includegraphics[width=\linewidth]{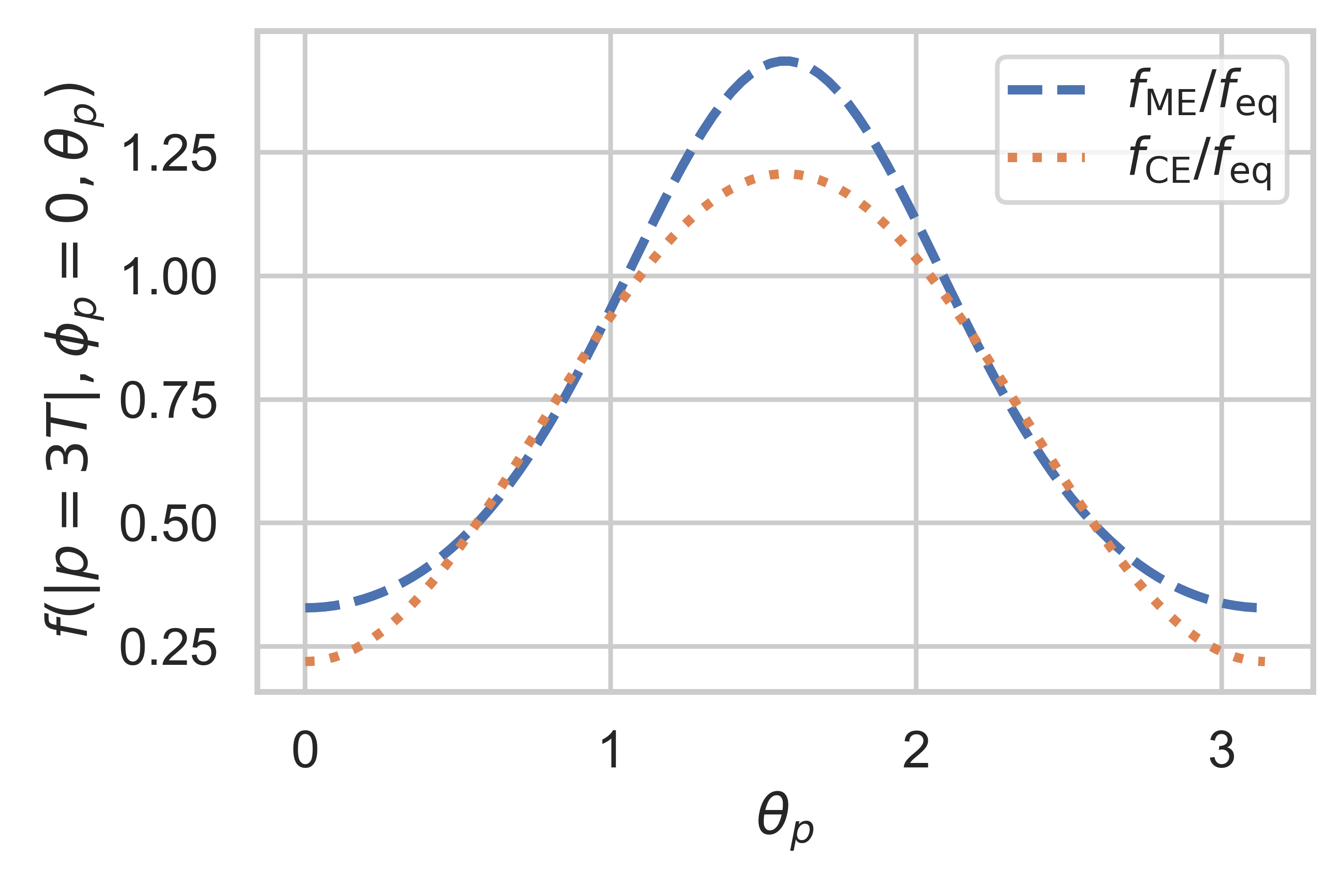}\\[-2ex]
\includegraphics[width=\linewidth]{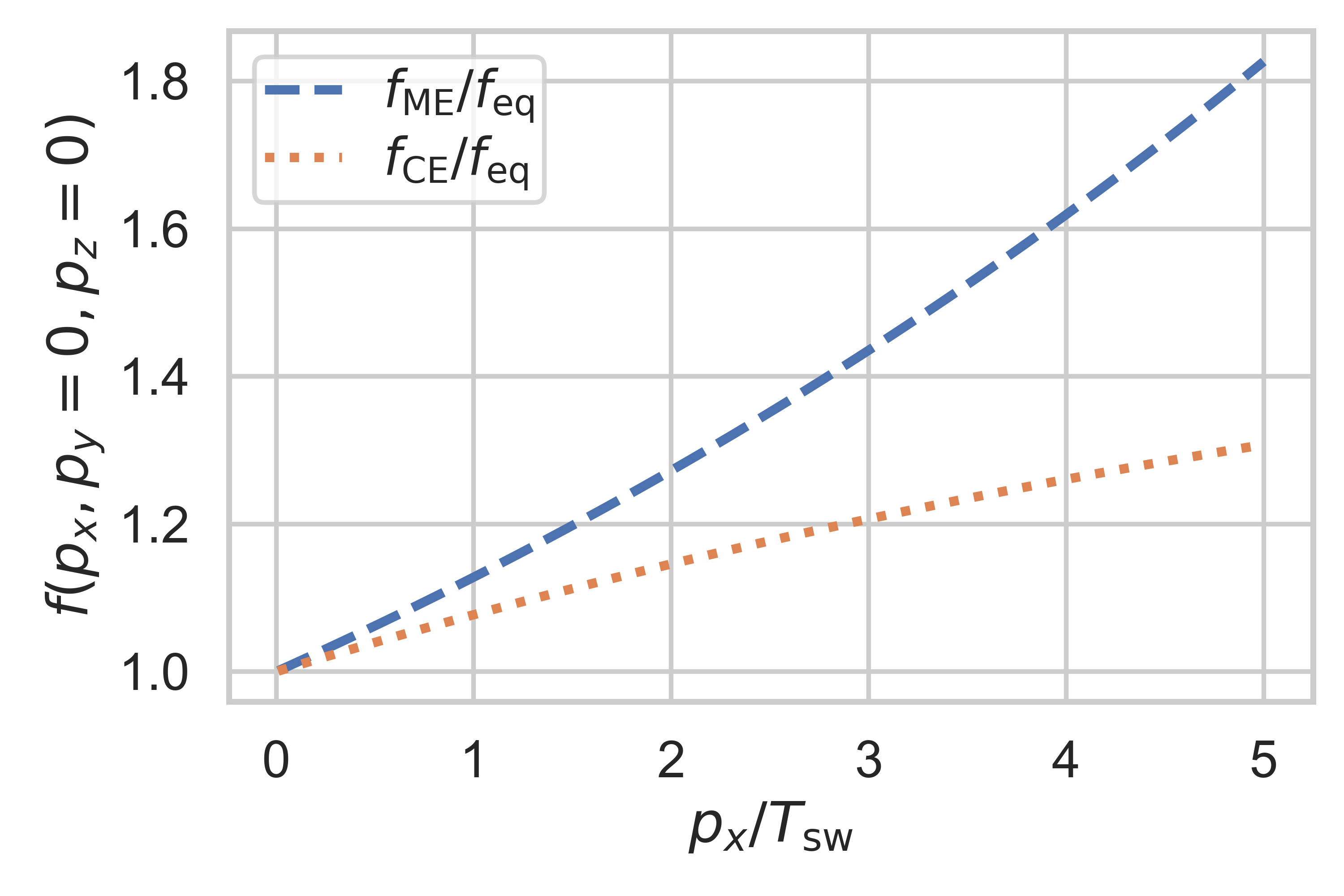}\\[-3ex]
\caption{
    The maximum-entropy (ME, blue dashed) and linear Chapman-Enskog RTA (CE, red dotted) distributions in the local rest frame as functions of polar momentum angle $\theta_p$ (top) and momentum magnitude (bottom) for a diagonal shear stress with $(x, y)$ isotropy, taking $\pi^{xx} = \pi^{yy} = p_{\rm eq}/5$. Both are normalized by the equilibrium distribution. 
    \vspace*{-5mm}}
  \label{fig:shear_thetap_px}
\end{figure}
%

In Figs.~\ref{fig:shear_phip} and \ref{fig:shear_thetap_px} we plot the momentum distribution in the local rest frame for two simple cases. In both cases, we take an energy density which corresponds (for a massless Boltzmann gas) to an equilibrium temperature of $T_{\rm sw} = 0.15$\,GeV.

In the first case, we assume the shear stress is isotropic in $(x,z)$ but anisotropic in $(x,y)$, taking $\pi^{xx} = \pi^{zz} = p_{\rm eq} / 5$ (where $p_{\rm eq}$ is the equilibrium pressure) and off-diagonal components zero. It follows that $\pi^{yy} = -\frac{2}{5}p_{\rm eq}$. Fig.~\ref{fig:shear_phip} shows the resulting variation of the maximum-entropy distribution as a function of the azimuthal angle $\phi_p$ for particles with momenta of average thermal magnitude $p = 3 T_{\rm sw}$. Also shown is the linear Chapman-Enskog RTA distribution, with viscous correction given by Eq.~(\ref{eqn:delta_f_CE_RTA}). 

In the second case, we take a shear stress tensor which is again diagonal but now isotropic in $(x, y)$, given by $\pi^{xx} = \pi^{yy} = p_{\rm eq} / 5$. It follows that $\pi^{zz} = -2 p_{\rm eq} / 5$. In Fig.~\ref{fig:shear_thetap_px} the maximum-entropy and linear Chapman-Enskog RTA distributions are plotted for particles with average thermal momentum $p=3T$ as a function of polar angle (left panel), and for fixed direction $\theta_p=\pi/2,\,\phi_p=0$ as functions of the momentum magnitude (right panel). 

\begin{figure}[!hb]
\includegraphics[width=0.98\linewidth]{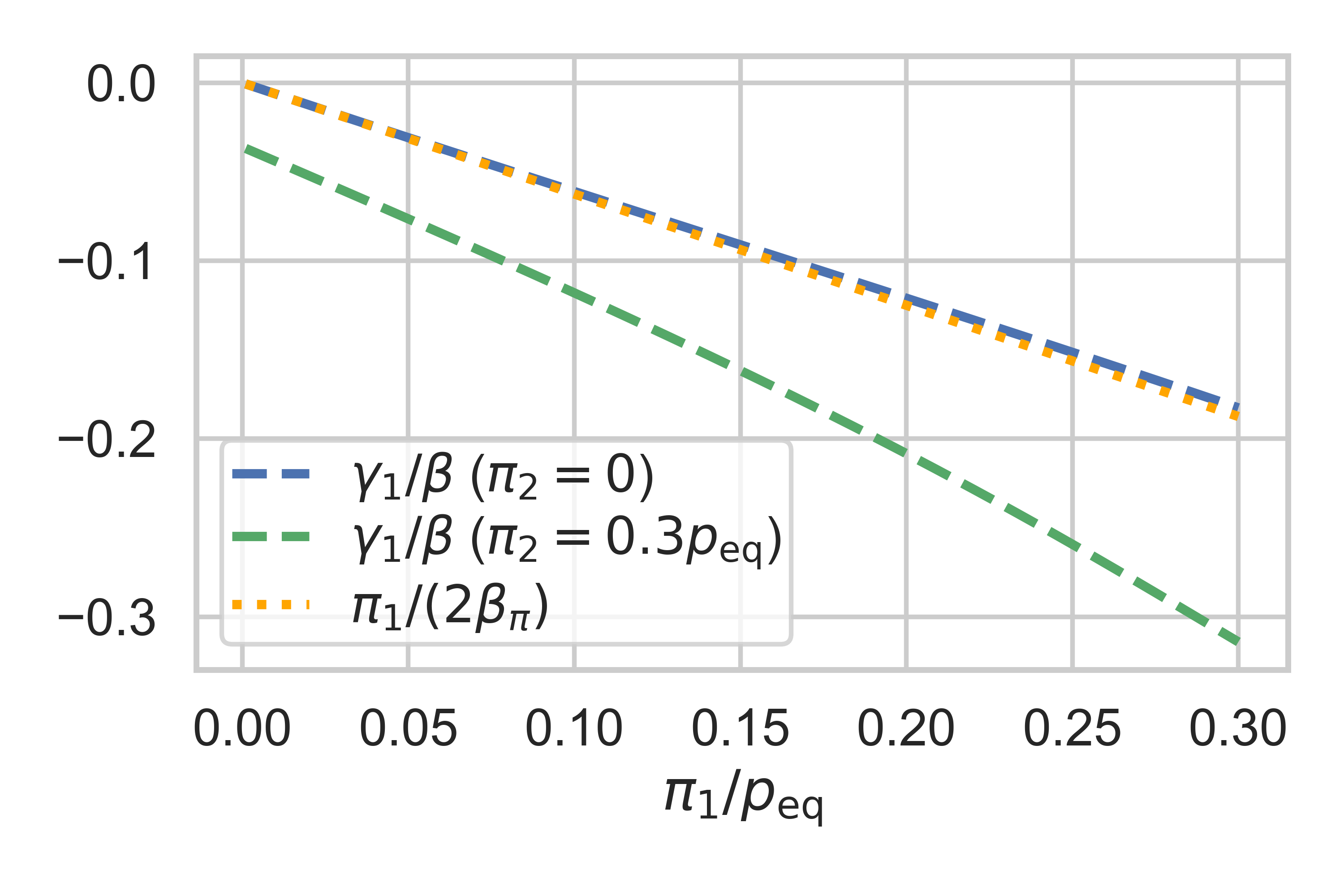}\\[-3ex]
\caption{
    The normalized Lagrange multiplier $\gamma_1/\beta$ (where $\beta$ is the inverse equilibrium temperature) as a function of the first shear stress eigenvalue $\pi_1$, for two choices of its second eigenvalue $\pi_2$, $\pi_2=0$ (blue dashed) and $\pi_2/p_{\rm eq} = 0.3$ (green dashed), respectively. Also shown for comparison is the corresponding CE RTA coefficient (orange dotted) which is independent of $\pi_2$. 
    }
\label{fig:gamma_1}
\end{figure}

Finally, in Fig.~\ref{fig:gamma_1} we plot the evolution of the Lagrange multiplier $\gamma_1$ (the first eigenvalue of $\bm{\gamma}$) with the associated shear stress eigenvalue $\pi_1/p_{\rm eq}$ (which is related to the inverse shear Reynolds number), for two choices of the second shear stress eigenvalue, $\pi_2/p_\mathrm{eq}= 0$ and 0.3, respectively. It is compared with the corresponding CE RTA coefficient which is linear in $\pi_1$ and independent of $\pi_2$. In both cases the energy density is fixed by the equilibrium energy density of the massless Maxwell-Boltzmann gas at a temperature $T = \beta^{-1} =0.15$ GeV. For $\pi_2=0$, the two coefficients agree very well, even for large inverse Reynolds numbers. However, agreement in the coefficients should not be interpreted as agreement in the predicted distributions: their functional forms are different (exponential momentum dependence in the ME distribution, polynomial dependence in the linearized RTA CE distribution). The green dashed line in Fig.~\ref{fig:gamma_1} illustrates that for non-zero $\pi_2$ the coefficients $\gamma_1/\beta$ and $\pi_1/(2\beta_\pi)$ differ and do not agree with each other even for very small $\pi_1$. This is a direct manifestation of the non-linear coupling between the eigenvalues $\gamma_1$ and $\gamma_2$ in the ME matching of the shear-stress. 

Although the method for finding the Lagrange multipliers $(\Lambda,\gamma_{\langle\mu\nu\rangle}$) from a given energy-momentum tensor with shear stress were here demonstrated numerically only for a massless Maxwell-Boltzmann gas, the framework for doing so for a general gas mixture of massive hadron resonances with Boltzmann statistics has been provided in this work, and its generalization to account in the generating function $\mathcal{Z}$ for Bose-Einstein or Fermi-Dirac statistics should be straightforward. However, a more efficient numerical routine for evaluating the momentum integrals in the massive particle case needs to be developed.

\section{Conclusions and outlook}

We have worked out the maximum-entropy distribution function as an alternative prescription for particlizing a fluid in a heavy-ion collision. For a general gas mixture of massive hadron resonances, we were able to solve numerically for the distribution function in the case that there was a non-zero bulk viscous pressure while the shear stress vanished. By comparing with the linear Chapman-Enskog RTA prescription we found that the maximum-entropy method yields significantly different particle momentum distributions and yields which can have non-negligible consequences for the theoretical interpretation of experimental data. For a gas of massless Maxwell-Boltzmann particles we demonstrated an algorithm for finding the maximum-entropy distribution when particlizing a fluid with vanishing bulk viscous pressure but non-zero shear stress. A full numerical solution of the maximum-entropy distribution for a massive hadron resonance gas in which both bulk and shear viscous stresses are nonzero is outstanding but of high value for phenomenological modeling of heavy-ion collisions.

Although in the present work we have not included any conserved charges such as net baryon number and strangeness, the generalization of the maximum-entropy prescription to include related dissipative effects (such as non-vanishing baryon and strangeness diffusion currents) should be straightforward. In general, this method allows to match the distribution function at particlization to {\it any} macroscopic quantity of which we have prior knowledge on the particlization surface.

Modern phenomenological studies of experimental heavy-ion collision data aim at reconstructing from the data, with quantified uncertainties, key parameters characterizing the evolving hot and dense medium (see, e.g., Refs.~\cite{Nijs:2020ors, Nijs:2020roc, Everett:2020yty, Everett:2020xug} for very recent examples of this type of approach). This is done within a Bayesian statistical framework in which the inferred probability distribution for the model parameters of interest (the ``posterior'') is obtained as the product of a ``prior'' distribution for the parameters (accounting for any prior knowledge that we might possess before performing the model-data comparison) and a ``likelihood'' which accounts for how well, for a given parameter choice, the model predictions agree with the measurements. 

An important consideration in Bayesian inference is to avoid introducing uncontrolled physics models in the likelihood that bias the parameter estimates. If assumptions made about the microscopic physics are not well-justified, the resulting model parameters won't be either. The maximum-entropy distribution introduced in this work provides a functional form for the unknown distribution of particles that implements {\it all of, and only} the information given to us by the hydrodynamic theory describing the dynamical evolution preceding the particlization process. In this sense it is {\it the least biased choice} that can be made in the absence of a trustworthy microscopic theory of the hadron gas close to the pseudo-critical temperature. Any other choice introduces additional information (``theoretical prejudice") into the particlization process that, as far as we know, cannot be compellingly justified theoretically.

\section*{Acknowledgments}

We thank Michael McNelis for very useful discussions regarding the comparisons between the maximum entropy and linear Chapman-Enskog RTA prescriptions. This work was supported by the National Science Foundation (NSF) within the framework of the JETSCAPE Collaboration under Award No. \rm{ACI-1550223}. Additional partial support by the U.S. Department of Energy (DOE), Office of Science, Office for Nuclear Physics under Award No. \rm{DE-SC0004286} and within the framework of the BEST and JET Collaborations is also acknowledged. 

\begin{appendix}
\end{appendix}

\bibliography{biblio}

\end{document}